\documentclass[12pt]{article}
\usepackage{amsmath,amssymb,amsthm,mathtools,bm}
\usepackage{graphicx,booktabs,threeparttable,caption,subcaption}
\usepackage{algorithm,algpseudocode,natbib}
\usepackage[dvipsnames,svgnames,x11names]{xcolor}
\usepackage[colorlinks=true,linkcolor=blue,citecolor=blue,urlcolor=blue,filecolor=Maroon]{hyperref}
\usepackage{placeins}
\graphicspath{{figures/}}
\newtheorem{theorem}{Theorem}
\newtheorem{proposition}{Proposition}
\newcommand{\R}{\mathbb{R}}
\newcommand{\E}{\mathbb{E}}
\newcommand{\Cov}{\operatorname{Cov}}

\newcommand{\med}{\operatorname{med}}

\newcommand{\Span}{\operatorname{span}}

\newcommand{\bhat}{\widehat{\boldsymbol\beta}}
\newcommand{\betaT}{\boldsymbol\beta}
\newcommand{\indep}{\perp\!\!\!\perp}
\newcommand{\anon}{1}
\begin{document}

\def\spacingset#1{\renewcommand{\baselinestretch}%
{#1}\small\normalsize} \spacingset{1}

\if1\anon
{
  \title{\bf Guarded Readmission Weighting for Robust Single-Index Fr\'echet Regression}
  \author{Fatma Sevin\c{c} Kurnaz\\
    Case Western Reserve University\\ Yildiz Technical University}
  \date{}
  \maketitle
} \fi

\if0\anon
{
  \bigskip
  \bigskip
  \bigskip
  \begin{center}
    {\LARGE\bf Guarded Readmission Weighting for Robust Single-Index Fr\'echet Regression}
  \end{center}
  \medskip
} \fi

\bigskip
\begin{abstract}
Single-index Fr\'echet regression reduces the predictor dimension for metric-space responses such as distributions, networks or curves, but its estimators are sensitive to outliers. We propose guarded readmission weighting (GRW), a one-step computation of case weights from the predictors and the response distances, usable with any weighted estimator. GRW starts from a high-breakdown fit to a multidimensional scaling surrogate, readmits cases by an add-one metric Cook's distance, downweights remote predictors and keeps a response isolation factor in the final weights. If fewer than $\lfloor n/2\rfloor$ of the $n$ cases are replaced, objects far from the uncontaminated responses receive weight zero; under elliptical predictors, weights depending on the predictors only through the index and the Mahalanobis distance keep the population weighted directions in the index subspace. In simulations, three GRW-weighted estimators keep the mean projection distance to the true direction below about 0.5 under up to 30\% scattered and 20\% clustered bad leverage points, whereas deletion and $\alpha$-distance covariance become much less accurate as contamination grows; $\alpha$-distance covariance is more accurate in most clean, shape-outlier and good-leverage settings. In two applications, under planted contamination, the weighted surrogate-assisted and least squares directions move less than those of methods without GRW.
\end{abstract}

\noindent%
\vfill

\newpage
\spacingset{1.8} 

\section{Introduction}\label{sec:intro}

Responses that are whole distributions now arise routinely. In the COVID-19 data analyzed by \citet{soale2025}, the response of each of the 254 Texas counties is the distribution of its reported seven-day case rates between August 2021 and February 2022; in the NHANES accelerometry data, the response of each participant is the distribution of minute-level physical activity over several days. Such objects cannot be inspected one by one, and a few unusual ones, such as small counties with few reports and many zero rates or participants with unusually high or low activity, can change the estimated relationship with the predictors. In the Texas data, when 10\% of the counties are replaced by cases with shifted predictors and responses, the direction estimated by surrogate-assisted ordinary least squares moves about as far as a direction drawn at random, with or without the deletion of flagged cases (Section~\ref{subsec:texas}).

Regression with responses that are random objects in a metric space --- distributions, networks, covariance matrices, curves --- and Euclidean predictors is now routinely analyzed through Fr\'echet regression \citep{petersen2019frechet}. In the single-index Fr\'echet model the response depends on the predictors only through one linear combination $\betaT^\top X$, which combines interpretability with nonparametric flexibility. The index can be estimated by M-estimation based on local Fr\'echet regression \citep{bhattacharjee2023single,ghosal2023frechet} or a neural network \citep{cui2026deep}, or by sufficient dimension reduction, in which inverse-regression and ordinary least squares (OLS) type estimators are lifted to object responses through kernels or surrogate responses built from the metric \citep{ying2022frechet,dong2022fksir,virta2022sir,zhang2024frechet,soale2023sdr}, with many later variants \citep{ying2025dwdr,huang2026knn,weng2024sparse,chen2026intrinsic,huang2026fdsdr}. These estimators use classical moments of the predictors together with all pairwise response distances or local averages, and, to our knowledge, none of them was designed for contaminated data or evaluated under contamination designs like those considered here; \citet{soale2026metric} note that some metrics are more sensitive to outliers than others. \citet{soale2025} addressed influential observations in the single-index model by deleting the cases flagged by a metric Cook's distance computed from a multidimensional scaling (MDS) coordinate of the responses. Like any one-at-a-time deletion rule, this is susceptible to masking when outliers come in groups, and it discards the information in cases that are only moderately influential.

For Euclidean responses the sensitivity of sliced inverse regression to outliers is well documented \citep{becker1998robust,prendergast2005influence,genschel2005robustness}, and robust versions target outliers in the predictors \citep{dong2015robust} or bound the influence function \citep{iaci2025robust}. Dependence-measure estimators reduce the effect of large distances by raising them to a power \citep{huang2024adcov} or through ball covariance \citep{zhang2019bcov,pan2020ball}; for distance covariance the influence function is bounded but the breakdown value is zero for every exponent \citep{leyder2025robust}, and the predictors are standardized by the classical covariance matrix, which a cluster of leverage points can distort. Robust Fr\'echet regression \citep{li2025robustfrechet,li2026robustlocal} and robust location and depth for objects \citep{kim2025frechetmedian,dai2023tukey,virta2023spatial} downweight or order individual responses but provide no dimension reduction. What is missing is protection of the \emph{direction estimate} against a fraction of cases contaminated in the response space, the predictor space or both, including clusters that mask themselves.

For this purpose we propose guarded readmission weighting (GRW), which computes case weights once from the predictors and the response distances. A high-breakdown start followed by one reweighting step is a classical pattern in robust regression \citep{coakley1993bounded,gervini2002class}; the contribution here is its construction for metric-space responses. Compared with the deletion of \citet{soale2025}, GRW differs in three respects. First, cases are judged against a high-breakdown start fitted to a weighted MDS surrogate. An add-one metric Cook's distance then compares cases inside and outside the start on the same scale, and the cases receive continuous weights instead of being deleted one at a time. Second, the predictor space is screened by a Mallows factor based on the minimum covariance determinant \citep{rvd1999}. Third, an isolation factor based on the median distance of each object to the others is kept in the final weights, so that an object far from all others cannot be readmitted because it is invisible in a low-dimensional MDS representation. Because GRW produces case weights, it can be used with any of the estimators above that accept case weights, and as a screen with those that do not; we study it with surrogate-assisted OLS (sa-OLS), Fr\'echet OLS (FOLS) and an OLS-type estimator on the weighted MDS surrogate, and as a screen for $\alpha$-distance covariance.

We compare GRW with the deletion of \citet{soale2025}, the only method designed for influential observations in single-index Fr\'echet regression, and with $\alpha$-distance covariance \citep[$\alpha$-dCov;][]{huang2024adcov}, the published robust sufficient dimension reduction method that comes closest to our setting and has public code \citep{rsdr2025}. This method obtains robustness through the dependence measure rather than through case weights and, because it standardizes the predictors by the classical covariance matrix, is not designed for leverage points. Since GRW produces weights, we also apply $\alpha$-dCov to the cases retained by GRW (GRW+$\alpha$-dCov). The single-index M-estimators are not included, because they require Fr\'echet means, which are not available for some of the metrics we use, and we found no public implementation.

We show that the add-one metric Cook's distance measures the effect of admitting a case relative to the robust start, that the positively weighted response objects stay in a bounded region when fewer than $\lfloor n/2\rfloor$ of the objects are replaced, and that, under elliptical predictors, weights that depend on the predictors only through the index and the Mahalanobis distance keep the population weighted directions in the index subspace. These results bound the inputs that receive positive weight and describe the weighting structure; they give no breakdown point or consistency result for the estimated direction. In simulations based on the models of \citet{soale2025}, GRW-weighted sa-OLS (GRW-SA) is more accurate than sa-OLS after deletion in 335 of 360 contaminated settings and, unlike $\alpha$-dCov, keeps the direction under bad leverage points up to 30\% contamination when they are scattered and up to 20\% when they are tightly clustered; $\alpha$-dCov is more accurate on clean data in most models and with shape outliers or good leverage points. In two applications, to the COVID-19 data of \citet{soale2025} and to physical-activity distributions from the National Health and Nutrition Examination Survey (NHANES), the directions of GRW-SA and of the GRW-weighted OLS estimator (GRW-OLS) move less than those of the methods that do not use GRW under planted bad leverage points and predict about as well as the other methods, but are more variable in clean data.

The paper makes three kinds of contribution. Methodologically, GRW replaces one-at-a-time deletion by continuous case weights, judges all cases on one scale through the add-one metric Cook's distance, and bounds the responses that receive positive weight. Computationally, the $n$ add-one Cook's distances are obtained in closed form from a single fit (Proposition~\ref{prop:cook}), the weights are computed once, and their cost is dominated by one eigendecomposition of the $n\times n$ distance matrix, which takes 3.5\,s for $n=1000$ (Supplement, Section~\ref{S-sec:timing}). Practically, the weights can be passed to any estimator that accepts case weights, their factors show why a case is downweighted, and R code with archives that reproduce every table and figure is provided.

Section~\ref{sec:model} introduces the weighted estimators, Section~\ref{sec:method} defines GRW, Section~\ref{sec:theory} gives the theoretical results, Section~\ref{sec:sims} the simulation studies and Section~\ref{sec:real} the applications; Section~\ref{sec:discussion} discusses limitations, and proofs and additional results are in the Supplement.

\section{Model and Preliminaries}\label{sec:model}
Let $X\in\R^p$ be a Euclidean predictor and let $Y$ take values in a metric space $(\Omega,d)$. We consider a single-index dimension-reduction model in which the conditional distribution of $Y$ given $X$ depends on the predictors only through the scalar index $T=\betaT^\top X$, where $\|\betaT\|=1$. Thus,
\begin{equation}Y\indep X\mid \betaT^\top X.\label{eq:si}\end{equation}
Because the index direction is identifiable only up to sign, our target is the one-dimensional subspace $\mathcal S=\Span(\betaT)$. The conditional-independence assumption in \eqref{eq:si} is stronger than the single-index assumption on the conditional Fr\'echet mean used by \citet{bhattacharjee2023single} and implies it whenever that mean exists.
For independent observations $(X_i,Y_i)$, $i=1,\ldots,n$, let $D_{ij}=d(Y_i,Y_j)$. Our construction builds on two established dimension-reduction approaches for metric-space responses: sa-OLS \citep{soale2023sdr} and FOLS \citep{ying2022frechet}. We first formulate their case-weighted counterparts for arbitrary nonnegative weights. The proposed method subsequently supplies these weights through a robust, data-adaptive procedure. The MDS coordinates used to assess the influence of observations are auxiliary quantities: they do not enter the final sa-OLS or FOLS estimators, but the first coordinate is the response of the GRW-weighted OLS estimator (Section~\ref{subsec:readmit}).

\subsection{Weighted Predictor Standardization}\label{subsec:standardization}
For nonnegative case weights $w_1,\ldots,w_n$ with $\sum_iw_i>0$, define $\pi_i=w_i/\sum_jw_j$ and
\begin{equation}\widehat\mu_w=\sum_{i=1}^n\pi_iX_i,\qquad \widehat\Sigma_w=\sum_{i=1}^n\pi_i(X_i-\widehat\mu_w)(X_i-\widehat\mu_w)^\top.\label{eq:weightedmoments}\end{equation}
If $\widehat\Sigma_w$ is positive definite, let $Z_i=\widehat\Sigma_w^{-1/2}(X_i-\widehat\mu_w)$. When $w_i=1$ for every $i$, these expressions reduce to the sample mean and the sample covariance matrix with divisor $n$. All subsequent weighted regression quantities are defined only when the required weighted covariance matrices are nonsingular.

\subsection{Weighted Surrogate-Assisted OLS}\label{subsec:sa}
Let $u_1,\ldots,u_N\in\R^n$ be unit-norm projection vectors, let $q>0$ denote a fixed distance power and set $E_{ij}=D_{ij}^q$. For $\ell=1,\ldots,N$, define the weighted distance surrogate and its weighted-centered counterpart by
\begin{equation}S_{i\ell}(w)=\sum_{j=1}^n\pi_jE_{ij}u_{j\ell},\qquad \widetilde S_{i\ell}(w)=S_{i\ell}(w)-\sum_{k=1}^n\pi_kS_{k\ell}(w).\label{eq:surrogates}\end{equation}
The standardized weighted regression directions and their candidate matrix are
\begin{equation}c_\ell(w)=\sum_{i=1}^n\pi_iZ_i\widetilde S_{i\ell}(w),\qquad \widehat M_{\mathrm{SA}}(w)=\frac1N\sum_{\ell=1}^Nc_\ell(w)c_\ell(w)^\top.\label{eq:msa}\end{equation}
Let $v_{\mathrm{SA}}$ be a leading unit eigenvector of $\widehat M_{\mathrm{SA}}(w)$. The resulting direction on the original predictor scale is $\bhat_{\mathrm{SA}}(w)\propto\widehat\Sigma_w^{-1/2}v_{\mathrm{SA}}$. The vectors $u_\ell$ are drawn independently of the data, uniformly from the unit sphere of $\R^n$; the population form of $\widehat M_{\mathrm{SA}}$ (Supplement, Section~\ref{S-sec:ellproof}) uses this distribution. The finite value of $N$ defines a Monte Carlo approximation to the projection-averaged candidate matrix; the distance power $q$ and the number of projections $N$ are specified in the numerical experiments.

\subsection{Weighted Fr\'echet OLS}\label{subsec:fols}
Here $E_{ij}=D_{ij}^q$ is as in Section~\ref{subsec:sa}, and the power $q$ need not be the same for the two estimators. Provided the denominator is positive and the resulting scale is nonzero, define
\begin{equation}\widehat\sigma_w^2=\frac{\sum_{i<j}\pi_i\pi_jE_{ij}^2}{\sum_{i<j}\pi_i\pi_j},\qquad \mathcal K_{ij}(w)=\exp\left\{-\frac{E_{ij}}{2\widehat\sigma_w^2}\right\}.\label{eq:bandwidth}\end{equation}
With $\widetilde{\mathcal K}_{ij}(w)=\mathcal K_{ij}(w)-\sum_{k=1}^n\pi_k\mathcal K_{kj}(w)$, define
\begin{equation}\kappa_j(w)=\sum_{i=1}^n\pi_iZ_i\widetilde{\mathcal K}_{ij}(w),\qquad \widehat M_{\mathrm{FOLS}}(w)=\sum_{j=1}^n\pi_j\kappa_j(w)\kappa_j(w)^\top.\label{eq:mfols}\end{equation}
If $v_{\mathrm{FOLS}}$ is a leading unit eigenvector of $\widehat M_{\mathrm{FOLS}}(w)$, the estimated direction is given by $\bhat_{\mathrm{FOLS}}(w)\propto\widehat\Sigma_w^{-1/2}v_{\mathrm{FOLS}}$. The weights thus change both the regression moments and, through the bandwidth, the kernel. The bandwidth \eqref{eq:bandwidth} is that of the FOLS code of \citet{soale2025}, which we keep for comparability with the results of that paper; because it is built from $E_{ij}^2$, the kernel is not invariant to a common rescaling $D\mapsto cD$ of the distances.

\subsection{Population Weighted Directions}\label{subsec:population}
For a measurable nonnegative weight function $w(X,Y)$ with $0<\E w<\infty$, define the tilted probability law $dP_w=w\,dP/\E w$, under which expectations and covariances are denoted by $\E_w$ and $\Cov_w$. For an admissible real-valued transformation $g$ of the response, let
\begin{equation}b_w(g)=\Sigma_w^{-1}\Cov_w\{X,g(Y)\},\qquad \Sigma_w=\Cov_w(X).\label{eq:bw}\end{equation}
Section~\ref{sec:theory} gives conditions under which these directions lie in $\mathcal S$; sample consistency, which would also require control of the estimated weights, the random projections or kernels and the leading eigenspace, is not established.

\section{Guarded Readmission Weighting}\label{sec:method}
Guarded readmission weighting (GRW) assigns continuous case weights that are then used in the weighted estimators of Section~\ref{sec:model}. It has five components: a response-space isolation factor, a robust start based on a weighted MDS surrogate, an add-one influence assessment, a Mallows-type predictor factor, and a final multiplicative guard. Readmission weighting (RW), the unguarded comparator, shares the start and the influence assessment but omits the final guard, so that comparing the two isolates the effect of the guard.

\subsection{Response-Space Isolation Factor}\label{subsec:guard}
Define the median response distances and their robust location and scale as
\begin{equation}o_i=\med_{j\ne i}D_{ij},\qquad m_o=\med_i o_i,\qquad s_o=1.4826\,\med_i|o_i-m_o|.\label{eq:iso}\end{equation}
For $s_o>0$, set $z_i=(o_i-m_o)/s_o$ and define $\psi(t)=[1-\{\min(1,\max(0,t))\}^2]^2$. The isolation factor is
\begin{equation}a_i=\psi\left(\frac{z_i-2}{2}\right),\qquad \tau_n=m_o+4s_o.\label{eq:guard}\end{equation}
Consequently, $a_i=1$ if $o_i\le m_o+2s_o$, $0<a_i<1$ if $m_o+2s_o<o_i<\tau_n$, and $a_i=0$ if $o_i\ge\tau_n$. Thus, observations with moderate isolation scores retain full isolation weight, whereas sufficiently isolated responses receive zero isolation weight. If $s_o=0$, which happens only when more than half of the $o_i$ coincide (for instance with
integer-valued distances), we use the limit of \eqref{eq:guard} as $s_o\downarrow0$: $a_i=1$ if
$o_i\le m_o$, $a_i=0$ otherwise, and $\tau_n=m_o$.

\subsection{Weighted MDS Surrogate and Robust Starting Fit}\label{subsec:start}
Let $\pi_i^{\mathrm{iso}}=a_i/\sum_ja_j$ and let $D^{\circ2}$ be the matrix with entries $d^2(Y_i,Y_k)$. Define $J_a=I_n-\mathbf1_n(\pi^{\mathrm{iso}})^\top$ and the weighted double-centered squared-distance matrix
\begin{equation}B_a=-\frac12J_aD^{\circ2}J_a^\top,\qquad G_a=W_a^{1/2}B_aW_a^{1/2},\quad W_a=\operatorname{diag}(a_1,\ldots,a_n).\label{eq:wmds}\end{equation}
Let $V_K$ contain the eigenvectors corresponding to the $K$ largest strictly positive eigenvalues of $G_a$, with diagonal eigenvalue matrix $\Lambda_K$. The weighted MDS surrogate scores, including the projections of observations with zero initial isolation weight, are
\begin{equation}S^{(a)}=B_aW_a^{1/2}V_K\Lambda_K^{-1/2}.\label{eq:mdsscores}\end{equation}
We take $K=\min\{2,K_+\}$, where $K_+$ denotes the number of eigenvalues of $G_a$ that exceed $10^{-8}$ times the largest one; the intended two-axis construction requires $K_+\ge2$, and if $K_+=1$ a single axis is used. The columns of $S^{(a)}$ are determined only up to sign; changing the sign of an axis changes neither $L_i$ in \eqref{eq:ltsresid} nor $\varrho_i$ in \eqref{eq:cookparts}. We assume that the $K$ leading eigenvalues are distinct and larger than the next one (here and for the matrix $G_{w^{(0)}}$ of Section~\ref{subsec:cook}), so that the axes are determined up to sign; a rotation within a tied eigenspace would leave $\varrho_i$ unchanged but not $L_i$, which is computed axis by axis. Algorithm~\ref{S-alg:grw} of the Supplement lists all steps, constants and numerical checks of GRW, including the orientation of the axes and the row orders that make the FAST-MCD and FAST-LTS computations below reproducible.

Let $(\widehat\mu_{\mathrm{MCD}},\widehat\Sigma_{\mathrm{MCD}})$ denote the reweighted minimum covariance determinant (MCD) location and scatter with coverage $h_{\mathrm{MCD}}=\lfloor(n+p+1)/2\rfloor$, computed by FAST-MCD \citep{rvd1999}, and define
\begin{equation}rd_i=\bigl\{(X_i-\widehat\mu_{\mathrm{MCD}})^\top\widehat\Sigma_{\mathrm{MCD}}^{-1}(X_i-\widehat\mu_{\mathrm{MCD}})\bigr\}^{1/2},\qquad q_x=\sqrt{\chi^2_{p,0.975}}.\label{eq:rd}\end{equation}
The robust start uses least trimmed squares (LTS) regression \citep{rousseeuw1984least}, computed by FAST-LTS \citep{rousseeuw2006computing}; its fitting subset is $\mathcal I_{\mathrm{MCD}}=\{i:rd_i<2q_x\}$. For each surrogate axis $k=1,\ldots,K$, let $(\widehat\alpha_k^{\mathrm{LTS}},\widehat b_k^{\mathrm{LTS}},\widehat s_k^{\mathrm{LTS}})$ denote an LTS regression fit of $S_{ik}^{(a)}$ on $(1,X_i^\top)$ over $i\in\mathcal I_{\mathrm{MCD}}$, with coverage $1/2$, and use the raw (unreweighted) LTS coefficients and scale. Define
\begin{equation}e_{ik}=\frac{S_{ik}^{(a)}-\widehat\alpha_k^{\mathrm{LTS}}-X_i^\top\widehat b_k^{\mathrm{LTS}}}{\widehat s_k^{\mathrm{LTS}}},\qquad L_i=\left(\sum_{k=1}^Ke_{ik}^2\right)^{1/2}.\label{eq:ltsresid}\end{equation}
The starting case weights are
\begin{equation}w_i^{(0)}=a_i\,\mathbf1\!\left\{L_i\le\sqrt{\chi^2_{K,0.975}}\right\}.\label{eq:startweights}\end{equation}
The GRW weights are defined only if the robust scatter matrix is nonsingular, the LTS scales are positive and $\sum_iw_i^{(0)}>p+1$, which implies that more than $p+1$ cases have positive start weight, and if the other conditions below \eqref{eq:addcook} hold; otherwise, in the simulations, the replication is recorded as a failure (Section~\ref{subsec:criteria}).

\subsection{Add-One Metric Cook's Distance}\label{subsec:cook}
Recompute the weighted MDS surrogate $S^{(0)}\in\R^{n\times K}$ by applying \eqref{eq:wmds}--\eqref{eq:mdsscores} with $w^{(0)}$ in place of $a$, keeping the selected value of $K$; this requires at least $K$ strictly positive eigenvalues of $G_{w^{(0)}}$. Let $\widetilde X=[\mathbf1_n,X]$, where $X$ here denotes the $n\times p$ matrix with rows $X_i^\top$, $W_0=\operatorname{diag}(w_1^{(0)},\ldots,w_n^{(0)})$, and define the weighted multivariate least squares fit and residuals by
\begin{equation}A_0=\widetilde X^\top W_0\widetilde X,\qquad \widehat B_0=A_0^{-1}\widetilde X^\top W_0S^{(0)},\qquad R_0=S^{(0)}-\widetilde X\widehat B_0.\label{eq:startfit}\end{equation}
Set $\widehat\Gamma_0=R_0^\top W_0R_0/(\sum_iw_i^{(0)}-p-1)$ and $\widetilde x_i=(1,X_i^\top)^\top$, and let $R_{0,i}$ denote the $i$th row of $R_0$, written as a column vector. For each observation, write
\begin{equation}h_i=\widetilde x_i^\top A_0^{-1}\widetilde x_i,\qquad d_i=1-w_i^{(0)}h_i,\qquad \widetilde h_i=h_i/d_i,\qquad \varrho_i=\frac{R_{0,i}^\top\widehat\Gamma_0^{-1}R_{0,i}}{K}.\label{eq:cookparts}\end{equation}
The add-one metric Cook's distance is defined as
\begin{equation}C_i=\frac{\widetilde h_i}{(1+\widetilde h_i)d_i^2(p+1)}\,\varrho_i.\label{eq:addcook}\end{equation}
The statistic $C_i$ evaluates the influence of admitting case $i$ relative to the robust starting fit: by Proposition~\ref{prop:cook} (Section~\ref{subsec:cookinterp}), it is the Cook-type distance between the weighted least squares fits in which case $i$ has weight zero and weight one, so that an initially excluded case with a small value of $C_i$ can be readmitted. The expression is defined when $A_0$ is nonsingular, $\sum_iw_i^{(0)}>p+1$, $\widehat\Gamma_0$ is positive definite and $d_i>0$. Since $A_0\succeq w_i^{(0)}\widetilde x_i\widetilde x_i^\top$, always $d_i\ge0$, and $d_i>0$ holds automatically when $w_i^{(0)}=0$. The numerical form of these checks in our implementation is given in Algorithm~\ref{S-alg:grw}. Unlike a deletion diagnostic based on refitting after removing each case, the present construction assesses addition relative to a common robust fit.

\subsection{Continuous Readmission and Final Guard}\label{subsec:readmit}
Let $n>p+1$, $c_0=4/(n-p-1)$, and retain the robust predictor distances $rd_i$ from \eqref{eq:rd}. The Cook-based readmission factor is defined as
\begin{equation}
r_i = \psi\left(\frac{C_i-c_0}{1.5c_0}\right).
\label{eq:cookweight}
\end{equation}

The Mallows-type predictor factor is given by
\begin{equation}
m_i^{X} = \psi\left(\frac{rd_i-3q_x}{q_x}\right).
\label{eq:mallowsweight}
\end{equation}

Combining these two factors yields the unguarded readmission weights,
\begin{equation}
w_i^{\mathrm{RW}} = r_i m_i^{X}.
\label{eq:rw}
\end{equation}

The proposed guarded readmission weights additionally incorporate
the response-space isolation factor $a_i$, giving
\begin{equation}
w_i^{\mathrm{GRW}} = a_i r_i m_i^{X}.
\label{eq:grw}
\end{equation}
The readmission factor attenuates observations with large add-one influence and the Mallows factor those with extreme predictor distances; the isolation factor in \eqref{eq:grw} limits the final weight of observations that are isolated in the response metric, including those whose influence is hidden in a low-dimensional MDS representation. The weighting is a single step; the readmission factor is continuous in $C_i$, whereas the start \eqref{eq:startweights} is a hard screen. Substituting $w^{\mathrm{GRW}}$ into \eqref{eq:msa} and \eqref{eq:mfols} defines GRW-SA and GRW-FOLS. The weighted least squares regression, with weights $w^{\mathrm{GRW}}$, of the first column of $S^{(0)}$ on the predictors defines GRW-OLS, an OLS-type estimator like sa-OLS and FOLS whose response surrogate is the first weighted MDS coordinate. RW-SA, RW-FOLS and RW-OLS use $w^{\mathrm{RW}}$ instead and serve as comparators in the ablation study (Section~\ref{subsec:ablation}). Finally, GRW+$\alpha$-dCov denotes $\alpha$-dCov (Section~\ref{subsec:methods}) applied to the cases with $w_i^{\mathrm{GRW}}>0$; it is a combination, not a weighted estimator.

\textbf{Invariance.} When the keys that order the rows for FAST-MCD and FAST-LTS (the row sums of $D$ and the surrogate scores; Algorithm~\ref{S-alg:grw}) have no ties, the GRW and RW weights do not change under affine transformations $X_i\mapsto AX_i+b$ of the predictors with nonsingular $A$ or under a rescaling $D\mapsto cD$ of the distances, and they are permuted with the cases (Supplement, Section~\ref{S-sec:invariance}). These statements hold in exact arithmetic; for the rescaling they also require that the surrogate scores, which are rounded to 10 decimal places before they are ordered, keep their order after rounding. In floating-point arithmetic the weights then agree up to rounding error, unless a quantity lies within rounding error of one of the cut-offs. Under an affine transformation, the GRW-weighted directions are proportional to $A^{-\top}\bhat$, so the estimated index does not depend on the units of the predictors. GRW-SA and GRW-OLS also do not depend on the scale of the distances, whereas GRW-FOLS does, through the bandwidth \eqref{eq:bandwidth}. Ties in the ordering keys can occur when the response distances are discrete, as for Poisson samples or networks; the weights can then depend on the order of the cases.

\section{Theoretical Properties}\label{sec:theory}
This section gives three results: an interpretation of the add-one metric Cook's distance, a bound on the responses that receive positive weight when part of the sample is replaced, and the behavior of weighted directions under elliptical predictors. The proofs are in the Supplement.

\subsection{Interpretation of the Add-One Metric Cook's Distance}\label{subsec:cookinterp}
For a case $i$, let $\widehat B_0^{(i,0)}$ and $\widehat B_0^{(i,1)}$ be the weighted least squares coefficients of the regression of $S^{(0)}$ on $\widetilde X$ in \eqref{eq:startfit} when the weight of case $i$ is set to zero and to one, respectively, all other weights being those of the start, and let $A_0^{(i,1)}=\widetilde X^\top W_0^{(i,1)}\widetilde X$, where $W_0^{(i,1)}$ is $W_0$ with its $i$th diagonal entry set to one.
\begin{proposition}\label{prop:cook}
If $A_0$ is nonsingular, $\widehat\Gamma_0$ is positive definite and $d_i>0$, then
\[
C_i=\frac{\operatorname{tr}\bigl\{(\widehat B_0^{(i,1)}-\widehat B_0^{(i,0)})^\top A_0^{(i,1)}(\widehat B_0^{(i,1)}-\widehat B_0^{(i,0)})\,\widehat\Gamma_0^{-1}\bigr\}}{K(p+1)}.
\]
\end{proposition}
\begin{proof}Supplement, Section~\ref{S-sec:cook}.\end{proof}
Thus $C_i$ is the change in the fitted coefficients caused by admitting case $i$ with full weight, measured, as in Cook's distance, in the metric $A_0^{(i,1)}$ of the fit that contains the case and scaled by the residual scatter $\widehat\Gamma_0$ of the start. For a case with $w_i^{(0)}=1$ it is the deletion Cook's distance of the weighted start fit, for a case with $w_i^{(0)}=0$ it measures the effect of adding the case, and for $0<w_i^{(0)}<1$, which occurs when $0<a_i<1$, it compares weights zero and one. Cases inside and outside the start are therefore judged on the same scale, and all $n$ values are obtained from one fit. Because the proof uses $\widehat\Gamma_0^{-1}$ only in quadratic forms, the identity also holds for the regularized inverse that our implementation uses when $\widehat\Gamma_0$ is nonzero but singular (Algorithm~\ref{S-alg:grw}).

\subsection{Bounded Responses under Replacement}\label{subsec:bounded}
Intuitively, an object that is far from more than half of the others has a large median distance $o_i$ and therefore receives isolation factor zero, and replacing fewer than $\lfloor n/2\rfloor$ of the cases cannot change this. Formally, in any sample, a case with $w_i^{\mathrm{GRW}}>0$ has $a_i>0$ and hence $o_i<\tau_n$ by \eqref{eq:guard} ($o_i\le\tau_n$ when $s_o=0$). Because $d(Y_i,Y_k)\le o_i+o_k$ for every pseudometric and $n\ge3$ (Supplement, Section~\ref{S-sec:radius}), the response objects of any two positively weighted cases are at distance less than $2\tau_n$ (at most $2\tau_n$ when $s_o=0$). The next result extends this bound to contaminated samples. Let $\mathcal Z_n=\{(X_i,Y_i)\}_{i=1}^n$ be a sample with response diameter $d_{\max}=\max_{i,k}d(Y_i,Y_k)>0$, let $\mathcal Z'_n$ be obtained by replacing $m$ of the observations by arbitrary pairs $(x,y)\in\R^p\times\Omega$, and let primes denote quantities computed from $\mathcal Z'_n$.
\begin{theorem}\label{thm:replace}
Let $d$ be a pseudometric, $n\ge3$ and $m\le\lfloor n/2\rfloor-1$, and assume that the GRW weights are defined on $\mathcal Z'_n$ (Sections~\ref{subsec:start} and~\ref{subsec:cook}). Then:
\begin{enumerate}
\item[(a)] $\tau'_n\le(1+4\times1.4826)\,d_{\max}<7d_{\max}$;
\item[(b)] every response object with a positive GRW weight, replaced or not, lies within distance $\tau'_n+d_{\max}<8d_{\max}$ of every response object that was not replaced, and any two positively weighted objects are at distance at most $2\tau'_n<14d_{\max}$;
\item[(c)] a case whose response object is at distance at least $8d_{\max}$ from some unreplaced object has $w^{\mathrm{GRW}}=0$; it does not enter GRW-SA, GRW-FOLS or GRW-OLS and can affect them only through the weights of the other cases.
\end{enumerate}
\end{theorem}
\begin{proof}Supplement, Section~\ref{S-sec:radius}.\end{proof}
The bounds do not depend on the replacing objects and follow from the isolation factor alone; what is specific to GRW is that this factor is kept in the final weights. Without it, an object that is far from all others but orthogonal to the two MDS axes is excluded from the start, but it can have a small surrogate residual and be readmitted with $w_i^{\mathrm{RW}}=1$; the model of the ablation study (Section~\ref{subsec:ablation}) is constructed so that this happens, and RW loses the direction there. Theorem~\ref{thm:replace} bounds the inputs that receive positive weight, not the angle of the estimated direction; a positive breakdown point of the direction remains open. It also requires the weights to be defined, which can fail under 30--40\% clustered leverage contamination (Section~\ref{subsec:simresults}).

\subsection{Weighted Directions under Elliptical Predictors}\label{subsec:elliptical}
Intuitively, under elliptical predictors the component of $X$ orthogonal to the index is symmetric given the index, its Mahalanobis distance and the response, so weights that depend only on these quantities cannot create a covariance between that component and the response. The GRW weight depends on the predictors through the residuals of the surrogate regressions, through the leverage and through the robust distance. The next result shows that, under elliptical predictors, weights that depend on the predictors only through $T$ and the Mahalanobis distance $Q$ keep the population weighted directions in the index subspace; whether the GRW weight has this form is discussed after the result.
\begin{theorem}\label{thm:ell}
Suppose that \eqref{eq:si} holds and that $X=\mu+\Sigma^{1/2}Z$ with $\Sigma$ positive definite, $\Sigma^{1/2}$ its symmetric square root, and $Z$ spherically symmetric with $\E\|Z\|<\infty$, and let $Q=(X-\mu)^\top\Sigma^{-1}(X-\mu)$. Let $w(X,Y)=\omega(T,Q,Y)$ for a measurable function $\omega$ with values in $[0,1]$, and suppose that $\E w>0$, $\E_w\|X\|^2<\infty$, and $\Sigma_w$ is nonsingular. Then $b_w(g)\in\Span(\betaT)$ for every measurable real function $g$ with $\E_w|g(Y)|<\infty$ and $\E_w\{\|X\|\,|g(Y)|\}<\infty$.
\end{theorem}
\begin{proof}Supplement, Section~\ref{S-sec:ellproof}.\end{proof}
Here $Z$ is the standardized predictor of the population model, not the $Z_i$ of Section~\ref{subsec:standardization}. The weighted estimators of Section~\ref{sec:model} estimate directions of this form: the weighted least squares slope of a response surrogate $f(Y)$ on $X$ is $b_w(f)$, and the population candidate matrices of weighted sa-OLS and FOLS are averages of outer products $b_w(g)b_w(g)^\top$ over transformations $g$ defined by the response distances (Supplement, Section~\ref{S-sec:ellproof}). Under the conditions of Theorem~\ref{thm:ell}, and when the required moments are finite, these matrices therefore have their range in $\Span(\betaT)$; when they are nonzero, the back-transformed leading eigenvector computed by the algorithms of Section~\ref{sec:model} spans $\Span(\betaT)$, and so does the slope when it is nonzero.

For weights that depend on the predictors only through $Q$, the conclusion is elementary, because such a reweighting preserves the spherical symmetry of $Z$; case weights that restore the conditions of inverse-regression methods were proposed by \citet{cook1994reweighting}. The point of Theorem~\ref{thm:ell} is that the weights may in addition depend on the index and on the response. Theorem~\ref{thm:ell} is a property of the weighting structure, not a consistency result for GRW. The GRW weight would have the stated form in the population if the robust distance, the leverage and the starting residuals depended on $x$ only through $T$ and $Q$; for the robust distance this holds when the MCD functional is unique, but for the residuals it requires unique LTS fits, which can fail, for example for links that are even in the index (Section~\ref{subsec:simresults}). We do not establish the population form of the GRW weight or the convergence of the sample weights and directions.

\section{Simulation Studies}\label{sec:sims}
We compare GRW with the classical estimators, with influential-observation deletion
\citep{soale2025} and with a published robust dimension-reduction method on the models of \citet{soale2025}, clean and contaminated in the response space, the predictor space or both (Sections~\ref{subsec:simmodels}--\ref{subsec:simresults}), and quantify the contribution of each component of GRW in an ablation study (Section~\ref{subsec:ablation}). All tuning constants of GRW are those of Section~\ref{sec:method} and are the same in every setting.

\subsection{Data-Generating Models}\label{subsec:simmodels}
We use the eight models of \citet{soale2025} and keep their numbering, summarized in
Table~\ref{tab:models}. In all of them the predictors are multivariate
$t$ with 20 degrees of freedom, $X=Z_0/\sqrt{U_0/20}$ with $Z_0\sim N_p(0,I_p)$ and $U_0\sim\chi^2_{20}$
independent, the index is $T=\betaT^\top X$ with $\betaT=(1,1,0,\ldots,0)^\top/\sqrt2$, and
$(n,p)\in\{(100,5),(200,10)\}$, so that two predictors are active and $p-2$ are inactive. Models I
and II have bivariate Euclidean responses, $Y=\{\sin(\pi/2+T),\sin(\pi/2-2T)\}^\top+0.5\xi$ and
$Y=0.8\,(T^3,-8T^3)^\top+0.5\xi$ with $\xi\sim N_2(0,I_2)$. In Models III and IV the response is a
distribution, observed through a sample of size 100 from the mixture $0.6N(T,1)+0.4N(-T,2^2)$ and
from the Poisson distribution with mean $\exp(T)$, respectively. Models V and VI have network
responses on 20 nodes: an Erd\H{o}s--R\'enyi graph with edge probability
$\operatorname{logit}^{-1}(\sin T)$, and a Poisson stochastic block model with block probabilities
$(0.4,0.3,0.3)$, mean matrix $15I_3+\lceil\exp(T)\rceil\mathbf 1_3\mathbf 1_3^\top$ and zero diagonal. In Models VII and VIII
the response is a curve observed at 30 time points, $\alpha(s)+2\sin(\pi s/2+T)+0.5\xi(s)$
and $\alpha(s)+\sin(\pi s/2+T)+\sin(\pi s/2-T)+0.5\xi(s)$, where $\alpha(s)=2\sin(\pi+\pi s/5)$ and the $\xi(s)$ are independent standard normal errors; the 30 time points are drawn from $U(0,10)$ and are shared
by the curves of a data set. As in \citet{soale2025}, distributions are compared by the Wasserstein
distances $W_1$ (sa-OLS and Cook's distance) and $W_2$ (FOLS), networks by the degree-centrality
distance (CD) and the graph diffusion distance (DD), and curves by the Euclidean distance of their
first 20 discrete Fourier coefficients (DFT). In Models I and VIII the link is an even function of
the index, so that, for symmetric $X$, $\E\{Xg(Y)\}=0$ for every transformation $g$ of the response; no
OLS-type estimator that is unweighted, or weighted with weights that are even in $x$, can recover
$\betaT$ there. These two models are
therefore used only without contamination and are reported separately.

\begin{table}[t]\centering\small
\caption{Simulation models of \citet{soale2025}. The index is $T=\boldsymbol\beta^\top X$; the last column lists the response distances used (Cook's distance and sa-OLS use the first, FOLS the second for distributions).}\label{tab:models}
\begin{tabular}{@{}clll@{}}\toprule
Model & Response & Dependence on $T$ & Distance\\\midrule
I & vector in $\R^2$ & sine (even in $T$) & Euclidean\\
II & vector in $\R^2$ & cubic & Euclidean\\
III & distribution (normal mixture) & location of the components & $W_1$, $W_2$\\
IV & distribution (Poisson) & mean $\exp(T)$ & $W_1$, $W_2$\\
V & network (Erd\H{o}s--R\'enyi) & edge probability & CD, DD\\
VI & network (stochastic block model) & edge intensity & CD, DD\\
VII & curve & phase & DFT\\
VIII & curve & phase (even in $T$) & DFT\\\bottomrule
\end{tabular}\end{table}

\subsection{Contamination}\label{subsec:contam}
In Models II--VII a fraction $\varepsilon\in\{0.05,0.10,0.20,0.30,0.40\}$ of the $n$ cases, chosen at random, is replaced. With the coordinate vectors $e_j$ of $\R^p$, the direction $e_\perp=(e_3+e_4)/\sqrt2$, which is orthogonal to $\betaT$ and lies in the span of two inactive predictors, a fresh draw $X^\ast$ from the predictor distribution and the extreme index value $t^\ast=3.5$, the six contamination types are:
\begin{center}\footnotesize
\begin{tabular}{@{}llll@{}}\toprule
Type & Predictor of an outlier & Response of an outlier & Purpose\\\midrule
T1 & unchanged & magnitude outlier & outliers in the response only\\
T2 & unchanged & shape outlier & outliers of typical size, different structure\\
T3 & $X+4\betaT$ & clean law at the new index & good leverage\\
T4 & $3e_\perp+X^\ast$ & magnitude outlier & moderate bad leverage\\
T5 & $6e_\perp+X^\ast$ & magnitude outlier & far bad leverage\\
T6 & $6e_\perp+0.1X^\ast$ & magnitude outlier & tightly clustered bad leverage\\\bottomrule
\end{tabular}\end{center}
Magnitude outliers are objects of the right kind at the wrong place: the clean response law at index $t^\ast$ for Models II, III, IV and VI, a dense Erd\H{o}s--R\'enyi graph with edge probability $0.95$ for Model V, and the clean curve shifted by $+4$ for Model VII. Shape outliers are objects of typical size with a different structure: the clean vector response rotated by $90^\circ$ (Model II), a sample from $N(0,5^2)$ (III), a negative binomial sample with mean $\exp(T)$ and size $0.5$ (IV), a two-block graph with within- and between-block probabilities $0.9$ and $0.1$ (V), and a curve with doubled frequency (VII). In Model VI, the shape outlier is a Poisson graph with mean $5+\lceil\exp(T)\rceil$ for every edge, close to the average entry $5.1+\lceil\exp(T)\rceil$ of the clean mean matrix at the same index but without block structure. In T4 and T5 the bad leverage points are scattered, because the fresh draw $X^\ast$ spreads them like the clean predictors; in T6 they are tightly clustered. Because the predictors are spherically symmetric, a rotation that fixes $\betaT$ maps $e_\perp$ to any other unit direction orthogonal to $\betaT$ and leaves the distribution of the clean data unchanged; since the classical, deletion, MM and GRW-based estimators are affine equivariant (for GRW, Supplement, Section~\ref{S-sec:invariance}), their distribution of $\Delta$ is the same for every such direction, and for $\alpha$-dCov, whose objective is affine invariant, only its random starting value can make a difference. Together with the clean settings this gives $16+360=376$ cells, each with 500 replications.

\subsection{Methods Compared}\label{subsec:methods}
We compare unweighted, deletion-based and GRW-weighted versions of the two base estimators of Section~\ref{sec:model}, $\alpha$-dCov, and a naive robust baseline. All methods are applied to the same data set in every replication, so that comparisons are paired.

\emph{Classical estimators.} sa-OLS \citep{soale2023sdr} uses $N=1000$ random projections and $q=1$ with each distance of Table~\ref{tab:models} other than $W_2$, and FOLS \citep{ying2022frechet} uses $q=2$ with $W_2$ for Models III and IV and with the distances of Table~\ref{tab:models} otherwise.

\emph{Deletion} \citep{soale2025}. The leading classical MDS coordinate of the response distances is regressed on the predictors, cases whose metric Cook's distance is at least $4/(n-p-1)$ are removed, and sa-OLS or FOLS is applied to the remaining cases. The published implementation (Del-pub), the supplementary code of \citet{soale2025}, recomputes the leading MDS coordinate after each deletion; since the sign of an MDS coordinate is arbitrary, the recomputed coordinate can be the mirror image of the full-sample one, which inflates the corresponding Cook's distances (Section~\ref{subsec:texas}). This concerns the implementation, not the deletion rule. The corrected implementation (Del), the code of the public repository accompanying the paper, computes the classical Cook's distance of the regression of the full-sample coordinate on the predictors and needs no recomputation; in the figures, Del (SA) and Del (FOLS) denote sa-OLS and FOLS after this deletion.

\emph{GRW.} The weights of Section~\ref{sec:method} are computed from the detection distance of each model, the first distance in Table~\ref{tab:models} (Euclidean, $W_1$, CD, DFT), which the Cook's distance of deletion also uses; for networks the same weights are used with CD and DD. They are used in weighted sa-OLS (GRW-SA) and weighted FOLS (GRW-FOLS). We also report GRW-OLS (Section~\ref{subsec:readmit}). For networks, the sa-OLS- and FOLS-type results in the tables and figures of the main text use CD; the results with DD are in Table~\ref{S-tab:S-DD} of the Supplement.

\emph{$\alpha$-distance covariance} \citep{huang2024adcov}. We use the authors' R package \citep{rsdr2025} with its default exponent $\alpha=0.5$, $d=1$ (one index) and its default optimizer, without tuning; the optimizer is local and is started once from a random value, so that its solution can depend on the computer (Supplement, Section~\ref{S-sec:impl}). Because the package accepts Euclidean responses only, each response is passed through a Euclidean representation of a distance of Table~\ref{tab:models} (Supplement, Section~\ref{S-sec:impl}); for networks this is the vector of node degrees, so that $\alpha$-dCov works with a different distance than the other methods. GRW+$\alpha$-dCov applies the same method to the cases with positive GRW weight (the package does not accept case weights).

\emph{A naive robust baseline.} MM denotes the MM-estimator \citep{yohai1987high} of the regression of the first classical MDS coordinate of the detection distance on the predictors. Least squares on this coordinate estimates $\Span(\betaT)$ under the linearity condition \citep{li1989regression}; replacing it by a robust regression is the simplest robustification, but it does not protect the MDS coordinate itself against outlying objects. It differs from GRW-OLS in that the MDS coordinate is classical rather than weighted and the GRW-weighted least squares fit is replaced by an MM regression.

\subsection{Performance Criteria}\label{subsec:criteria}
For unit vectors $\bhat$ and $\betaT$ let $P_{\bhat}=\bhat\bhat^\top$ and $P_{\betaT}=\betaT\betaT^\top$. Accuracy is measured by the projection distance, the measure of breakdown used by \citet{genschel2005robustness},
\begin{equation}\Delta(\bhat,\betaT)=\|P_{\bhat}-P_{\betaT}\|_F=\sqrt{2}\,|\sin\theta|\in[0,\sqrt{2}],\label{eq:delta}\end{equation}
where $\theta$ is the angle between the directions; $\Delta=\sqrt2$ means that the estimated direction is orthogonal to $\betaT$. We report the mean of $\Delta$ over replications with its Monte Carlo standard error and paired differences between methods. A replication in which a method cannot be computed --- for GRW, when the weights are undefined (Section~\ref{subsec:simresults}) --- is recorded as a failure and reported separately, not discarded and replaced; in averages and paired comparisons it counts as $\Delta=\sqrt2$, the value of a direction orthogonal to $\betaT$. Secondary outcomes are the proportion of regular (uncontaminated) cases that receive weight zero, the retention of good leverage points in T3, and computing time; the replications of the main study took about 97 hours of computing time in total. On one core of a laptop the GRW weights take 3.5\,s for $n=1000$ and 27\,s for $n=2000$, whereas the published deletion algorithm of \citet{soale2025} takes 42\,s already for $n=500$ (Supplement, Section~\ref{S-sec:timing}).

\subsection{Clean Data and Contamination}\label{subsec:simresults}

\begin{table}[t]\centering\scriptsize\setlength{\tabcolsep}{2.6pt}
\caption{Mean $\Delta$ by contamination type, averaged over Models II--VII and, for T1--T6, over the contamination fractions 0.05--0.40 (500 replications per setting; a replication in which GRW cannot be computed counts as $\Delta=\sqrt2$).}\label{tab:summary}
\begin{tabular}{@{}l*{12}{c}@{}}\toprule
& \multicolumn{4}{c}{sa-OLS family} & \multicolumn{4}{c}{FOLS family} & \multicolumn{4}{c}{other}\\\cmidrule(lr){2-5}\cmidrule(lr){6-9}\cmidrule(l){10-13}
Type & classical & Del-pub & Del & GRW & classical & Del-pub & Del & GRW & GRW-OLS & MM & \(\alpha\)-dCov & \shortstack{GRW+\\\(\alpha\)-dCov}\\
\midrule
\multicolumn{13}{@{}l}{\emph{$(n,p)=(100,5)$}}\\
clean & 0.304 & 0.191 & 0.191 & 0.131 & 0.326 & 0.188 & 0.186 & 0.124 & 0.113 & 0.145 & \textbf{0.099} & 0.102\\
T1 & 0.880 & 0.762 & 0.729 & 0.169 & 0.900 & 0.786 & 0.747 & 0.164 & 0.141 & 0.323 & 0.384 & \textbf{0.126}\\
T2 & 0.474 & 0.360 & 0.350 & 0.159 & 0.530 & 0.383 & 0.367 & 0.153 & 0.142 & 0.357 & 0.143 & \textbf{0.121}\\
T3 & 0.596 & 0.499 & 0.491 & 0.166 & 0.680 & 0.544 & 0.538 & 0.160 & 0.139 & 0.273 & 0.131 & \textbf{0.121}\\
T4 & 1.277 & 1.005 & 1.010 & 0.188 & 1.288 & 1.024 & 1.035 & 0.184 & 0.157 & 0.385 & 0.987 & \textbf{0.145}\\
T5 & 1.269 & 0.943 & 1.088 & 0.218 & 1.271 & 0.951 & 1.096 & 0.214 & 0.188 & 0.393 & 1.040 & \textbf{0.182}\\
T6 & 1.242 & 0.889 & 1.059 & 0.463 & 1.242 & 0.887 & 1.052 & 0.452 & 0.440 & 0.484 & 1.016 & \textbf{0.429}\\
\multicolumn{13}{@{}l}{\emph{$(n,p)=(200,10)$}}\\
clean & 0.372 & 0.217 & 0.213 & 0.146 & 0.407 & 0.213 & 0.209 & 0.136 & 0.124 & 0.170 & \textbf{0.101} & 0.104\\
T1 & 0.982 & 0.834 & 0.812 & 0.186 & 1.002 & 0.855 & 0.833 & 0.180 & 0.155 & 0.344 & 0.430 & \textbf{0.127}\\
T2 & 0.566 & 0.409 & 0.400 & 0.175 & 0.641 & 0.424 & 0.411 & 0.165 & 0.155 & 0.395 & 0.146 & \textbf{0.121}\\
T3 & 0.664 & 0.551 & 0.544 & 0.180 & 0.779 & 0.609 & 0.604 & 0.172 & 0.150 & 0.293 & 0.151 & \textbf{0.123}\\
T4 & 1.294 & 1.070 & 1.079 & 0.209 & 1.300 & 1.093 & 1.099 & 0.204 & 0.173 & 0.452 & 1.004 & \textbf{0.152}\\
T5 & 1.282 & 1.099 & 1.154 & 0.251 & 1.281 & 1.109 & 1.168 & 0.246 & 0.217 & 0.467 & 1.053 & \textbf{0.201}\\
T6 & 1.254 & 1.094 & 1.115 & 0.457 & 1.251 & 1.087 & 1.111 & 0.446 & 0.433 & 0.517 & 1.016 & \textbf{0.424}\\
\bottomrule\end{tabular}
\par\smallskip\parbox{\linewidth}{\scriptsize Classical: unweighted sa-OLS or FOLS; Del-pub: deletion with the published code of Soale (2025); Del: deletion with the classical Cook's distance of the full-sample coordinate (corrected implementation); GRW: GRW-SA in the sa-OLS family and GRW-FOLS in the FOLS family; MM: MM-regression of the first MDS coordinate; GRW+$\alpha$-dCov: $\alpha$-dCov on the cases with positive GRW weight. Bold: smallest value in the row.}
\end{table}

\textbf{Clean data.} Weighting improves the OLS-type estimators even without contamination (Table~\ref{tab:summary}; per model in Table~\ref{S-tab:clean} of the Supplement): averaged over Models II--VII with $(n,p)=(100,5)$, mean $\Delta$ is 0.304 for sa-OLS, 0.191 after deletion and 0.131 for GRW-SA, and the advantage remains with normal predictors (Supplement, Section~\ref{S-sec:sens}). Since deletion also helps, part of the gain appears to come from limiting the few extreme responses that the links of these models produce even without contamination. The price is that GRW gives weight zero to 10--32\% of the cases (20.5\% on average). $\alpha$-dCov is the most accurate method on clean data in Models II, III, IV and VI, and GRW-OLS in Models V (tied with GRW-FOLS to three decimals) and VII; GRW+$\alpha$-dCov costs almost nothing (0.102 against 0.099). In Model VI with the diffusion distance, GRW is slightly less accurate than deletion (sa-OLS: 0.210 against 0.198; FOLS: 0.182 against 0.168; $n=100$; Table~\ref{S-tab:S-DD}).

\begin{figure}[p]\centering
\includegraphics[width=0.96\linewidth]{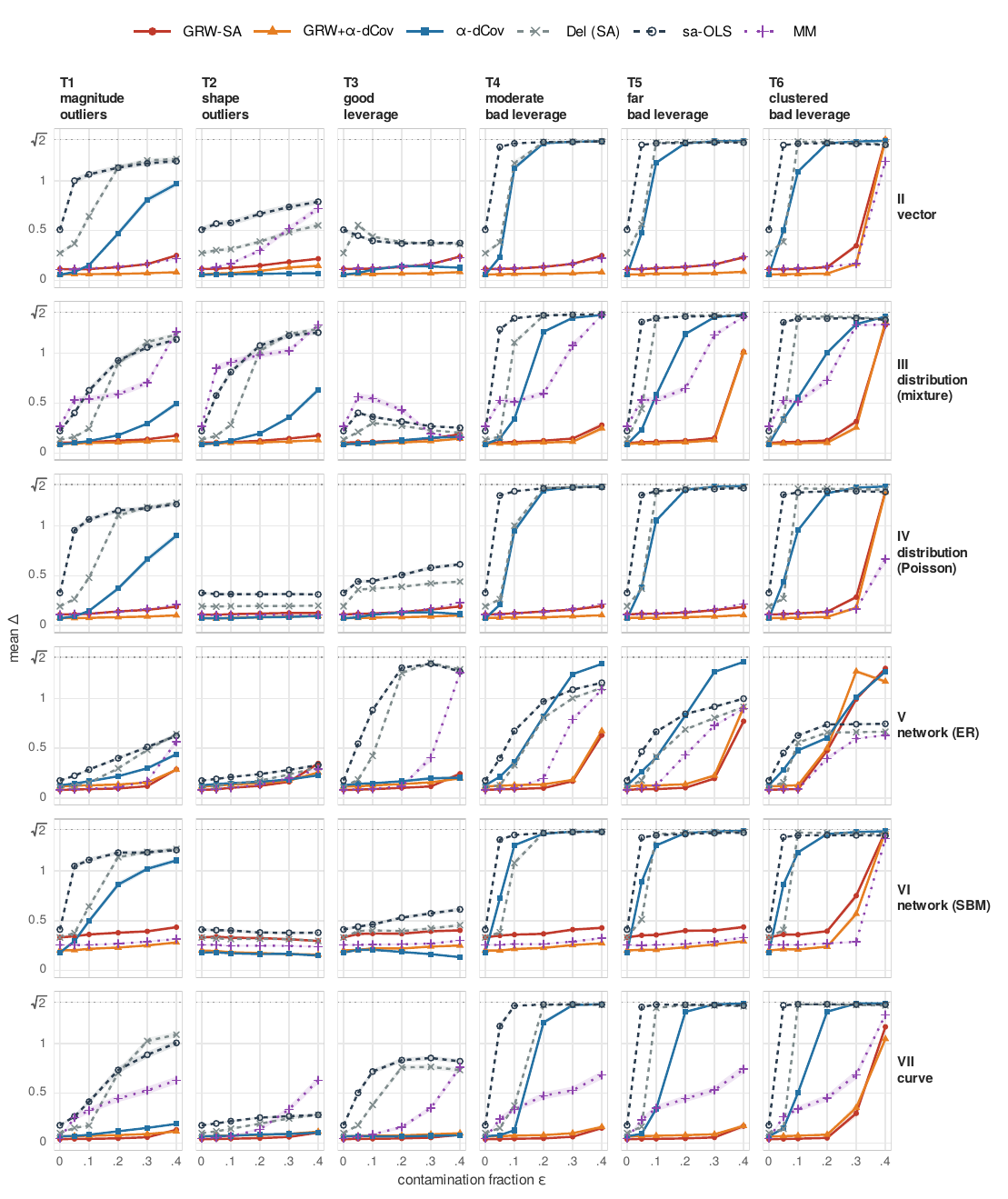}
\caption{Mean $\Delta$ against the contamination fraction, $(n,p)=(100,5)$, 500 replications; the value at $\varepsilon=0$ is the clean setting, bands are $\pm2$ Monte Carlo standard errors, and the dotted line is $\sqrt2$ (direction orthogonal to $\betaT$). sa-OLS-type methods use the detection distance (Euclidean, $W_1$, CD, DFT); sa-OLS is unweighted, Del (SA) is sa-OLS after the corrected deletion of \citet{soale2025}, GRW+$\alpha$-dCov is $\alpha$-dCov on the cases with positive GRW weight, and MM is the naive robust baseline of Section~\ref{subsec:methods}. ER: Erd\H{o}s--R\'enyi graph; SBM: stochastic block model. The results for $(n,p)=(200,10)$ are similar (Table~\ref{tab:summary}).}
\label{fig:delta}
\end{figure}

\textbf{Contamination.} The counts below are paired comparisons on the same data sets with differences larger than two Monte Carlo standard errors. Deletion is strongly affected by every contamination type, least by shape outliers (T2): at 20\% contamination and $(n,p)=(100,5)$ its mean $\Delta$ is between 0.38 (T2) and 1.29 (T4), whereas that of GRW-SA is 0.15--0.16 for T1--T5 and 0.22 for T6 (Figure~\ref{fig:delta}). GRW-SA is more accurate than deletion in 335 of the 360 contaminated settings and less accurate in 15, all with shape outliers in Model VI (5--20\%) and Model V (40\%) or with 30--40\% clustered leverage points (T6), where both are inaccurate; the other 10 differences are within two standard errors. The published implementation of deletion is more accurate than the corrected one under T5 and T6 (mean $\Delta$ 0.89--1.11 against 1.05--1.17), although both are far less accurate than GRW; for the other types, their mean $\Delta$, averaged over the models and contamination fractions, differs by at most 0.04, and the published code could not be computed in up to 1.2\% of the replications of a setting. Of the three GRW estimators, GRW-OLS has the smallest point estimate of mean $\Delta$ in 294 of the 372 settings of Models II--VII (360 contaminated and 12 clean).

\begin{figure}[t]\centering
\includegraphics[width=\linewidth]{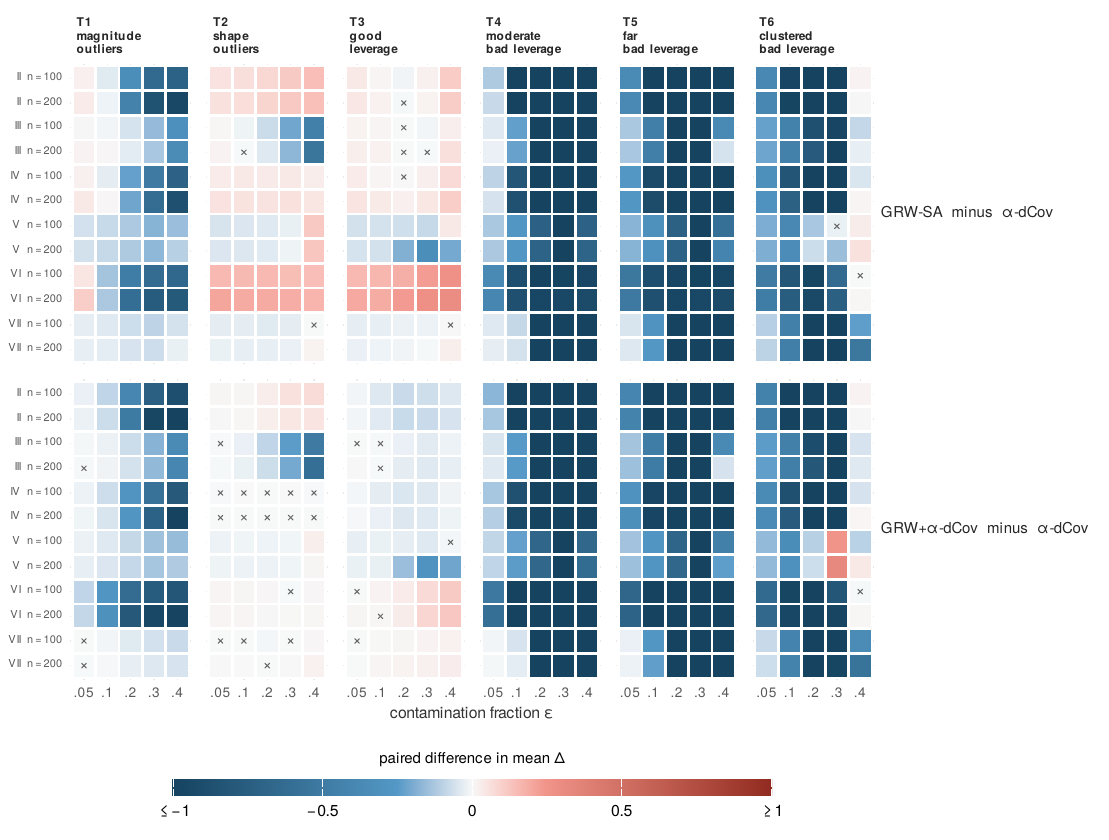}
\caption{Paired differences in mean $\Delta$ to $\alpha$-dCov, for every contaminated
setting (500 replications). Blue: the GRW-based method is more accurate; red: $\alpha$-distance
covariance is more accurate; a cross marks a difference smaller than two standard errors.}
\label{fig:paired}
\end{figure}

$\alpha$-dCov resists outliers that are not leverage points but breaks down under bad leverage: for T4 and T5 its mean $\Delta$, averaged over Models II--VII ($n=100$), grows from 0.27--0.39 at 5\% to more than 1.2 from 20\% contamination on, and GRW-SA is more accurate in all 120 of these settings (both sample sizes; Figure~\ref{fig:paired}) and in 50 of the 60 settings of T1. With shape outliers (T2) and good leverage points (T3), $\alpha$-dCov is more accurate in 35 of 60 settings each, mainly in Models II, IV and VI. GRW+$\alpha$-dCov has the smallest average $\Delta$ in every contamination type; it is more accurate than $\alpha$-dCov in all settings of T4 and T5 and in 57 of 60 of T1, and less accurate in 26 settings of T2 (by at most 0.08), 18 of T3 (by at most 0.12) and 7 of T6 (by at most 0.32, in Model V with 30\% contamination).

\textbf{Good leverage points.} GRW treats good leverage points like bad ones: it gives a weight above 0.5 to only 2\% of them (at most 18\% in any setting). In 20 data sets of each of Models II, IV, V and VII with 20\% good leverage points ($n=100$), at most 2\% of them receive a positive start weight, and $r_i>0.5$ for at most 2.3\% of them, against 73--90\% of the other cases, because the add-one metric Cook's distance grows with the leverage of a case. The Mallows factor is not the cause (Section~\ref{subsec:ablation}).

\textbf{Clustered leverage and failures.} Under tightly clustered bad leverage (T6), GRW-SA has mean $\Delta$ 0.14 at 10\%, 0.22 at 20\% (Model V: 0.48) and 0.50 at 30\% contamination ($n=100$); at 40\% all methods are inaccurate in most models. In Model V with 30--40\% T6 contamination deletion is more accurate than GRW, although both are inaccurate (at 40\%: 0.67 against 1.30). With 40\% T6 contamination the GRW weights cannot be computed in 88--100\% of the replications of Models II, IV and VI, mostly because the robust start leaves at most $p+1$ cases with positive weight; in the other settings with 30--40\% contamination this happens in at most 12\% of the replications (6.2\% at 30\%), and outside T6 in one replication out of 500 (Supplement, Section~\ref{S-sec:failures}). Failures count as $\Delta=\sqrt2$.

\textbf{Links that are even in the index.} In Models I and VIII (Supplement, Table~\ref{S-tab:even}) the unweighted OLS-type estimators have mean $\Delta$ 1.18 and 1.08 for sa-OLS and 1.17 and 1.06 for FOLS ($n=100$), close to the value of about 1.25 for a direction drawn uniformly at random when $p=5$. GRW-SA reduces mean $\Delta$ to 0.77 and 0.46, possibly because a coverage-$1/2$ LTS fit can follow one branch of the link, so that the weights are not even in $x$; but only $\alpha$-dCov recovers the direction (0.16 and 0.11), and GRW does not remove this limitation of the OLS-type estimators.

\textbf{Predictor distribution and dimension.} With normal and with skewed, nonelliptical predictors, $(n,p)=(100,5)$, and with $(n,p)=(200,20)$ in Models II, IV and VII (Supplement, Section~\ref{S-sec:sens}), the conclusions do not change: for each design, GRW-SA is more accurate than deletion in all 30 contaminated settings and than $\alpha$-dCov in all settings of T1, T5 and T6, its mean $\Delta$ averaged over the contaminated settings is the same with skewed and with normal predictors (0.099), and the robust start never failed.

\subsection{Ablation: Contribution of the Components of GRW}\label{subsec:ablation}
The ablation study uses Models II--VII with $(n,p)=(100,5)$, clean and with contamination of types T1, T2, T3, T5 and T6 at 10\% and 20\%, together with a model constructed to expose the failure mode that motivated the guard (Model O, for orthogonal outliers). In this model the response is $Y=(T+0.3\xi_1,\, 0.5T^2+0.3\xi_2,\,0.1\xi_3)\in\R^3$ with independent standard normal $\xi_j$ and Euclidean distance, so that the signal lies in the first two MDS axes; a fraction $\varepsilon\in\{0.01,0.05,0.10\}$ of the cases is replaced by outliers with predictor $2e_\perp$ plus a fresh draw and response $(0,0,20+0.1\xi_4)$, far from all other objects but orthogonal to the two MDS axes. Each cell has 200 replications. Each variant changes one component of GRW: (i) no final guard (RW, the unguarded procedure); (ii) classical instead of weighted MDS; (iii) no LTS start; (iv) no isolation factor, in the start and in the guard; (v) no Mallows factor (in our implementation this variant also fits the LTS start to all cases instead of $\mathcal I_{\mathrm{MCD}}$); (vi) the start weights only, without readmission; (vii) a hard 0/1 readmission cut-off; (viii) an additional factor for the orthogonal distance of an object to the two MDS axes (orthogonal-distance factor); and (ix) iterated instead of one-step reweighting. For each variant, $\Delta$ is the average of the $\Delta$ of its weighted sa-OLS and weighted FOLS directions, so that in this subsection GRW stands for the mean of GRW-SA and GRW-FOLS and RW for the mean of RW-SA and RW-FOLS. We report paired differences to GRW with their standard errors.

\begin{figure}[t]\centering
\includegraphics[width=0.96\linewidth]{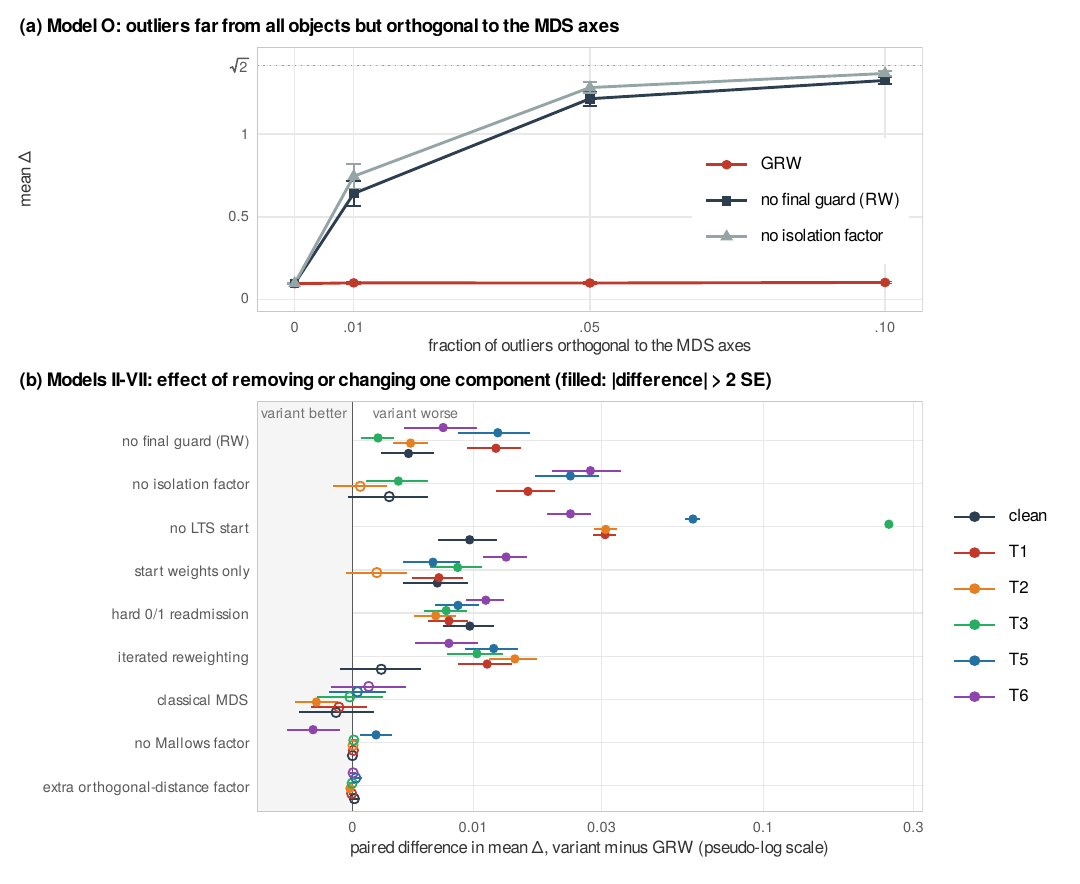}
\caption{Ablation study. $\Delta$ is the average for the weighted sa-OLS and FOLS directions of each variant. (a) Model O: mean $\Delta$ ($\pm2$ standard errors, 200 replications) of GRW, of the unguarded procedure RW and of GRW without the isolation factor. (b) Mean paired difference in $\Delta$ between each variant and GRW, by contamination type, Models II--VII, $(n,p)=(100,5)$; bars are $\pm2$ standard errors (SE) and filled symbols differ from zero by more than two standard errors.}\label{fig:ablation}
\end{figure}

Figure~\ref{fig:ablation} and Table~\ref{S-tab:ablation} of the Supplement summarize the results. In Model O, without the final guard one outlier in a hundred increases mean $\Delta$ from 0.10 to 0.64, and with 5\% such outliers the direction is lost (1.21 against 0.10). In Models II--VII the guard improves the accuracy slightly but consistently. Below, a/b gives the numbers of settings in which a variant is worse/better than GRW by more than two standard errors. Counting all 70 settings of the ablation study (Models II--VII and Model O), RW is worse than GRW in 44 settings and better in 2 (44/2). The LTS start matters most for good leverage points (T3; a difference of 0.25, almost entirely from Models V and VII) and for far bad leverage (T5; 0.06), and the isolation factor for T1, T5 and T6. Continuous readmission is better than a hard 0/1 cut-off (49/0), than the start weights alone (29/7) and than iterated reweighting (29/6). Classical instead of weighted MDS (10/8), removing the Mallows factor (3/1) and the additional orthogonal-distance factor (3/16) change mean $\Delta$ by less than 0.03 in every setting. We keep the Mallows factor because it gives weight zero to cases with very remote predictors ($rd_i\ge4q_x$), and the weighted MDS because it removes the influence of isolated objects on the surrogate axes; Theorem~\ref{thm:replace} does not depend on either choice. Averaged over Models II, IV and VII, changing any one tuning constant changes mean $\Delta$ by at most 0.005, except one surrogate axis instead of two and an LTS coverage of 0.75 instead of 1/2, which are worse by up to 0.022 and 0.014 (Supplement, Section~\ref{S-sec:tuning}).

\section{Real-Data Applications}\label{sec:real}
\subsection{COVID-19 Transmission in Texas}\label{subsec:texas}

We use the data of the first application of \citet{soale2025}. The responses are the distributions of the total new COVID-19 cases per 100{,}000 persons in the last seven days, reported at the county level between 1 August 2021 and 21 February 2022 for the $n=254$ counties of Texas \citep{cdc2023covid}; the predictors are $p=8$ demographic characteristics of the counties from the 2020 American Community Survey \citep{census2022acs}; see \citet{soale2023sdr} for a detailed description. As in \citet{soale2025}, sa-OLS uses the Wasserstein distance $W_1$ and FOLS uses $W_2$; the GRW weights are computed from $W_1$, and all methods of Section~\ref{subsec:methods} are applied with the settings of the simulations (details in Section~\ref{S-sec:texas} of the Supplement). These data fit the problem setting: the responses are distributions compared in a Wasserstein metric, a single index of the county characteristics is the quantity of interest, and counties with few reports produce responses that are far from the others (16 of the 45 counties with GRW weight zero have a median reported rate of zero; Supplement, Section~\ref{S-sec:texas}). Ellipticity of the predictors, which Theorem~\ref{thm:ell} assumes, is not checked here; in the simulation models, skewed predictors did not reduce the accuracy of GRW (Supplement, Section~\ref{S-sec:sens}).

\begin{figure}[!htbp]\centering
\includegraphics[width=0.88\linewidth]{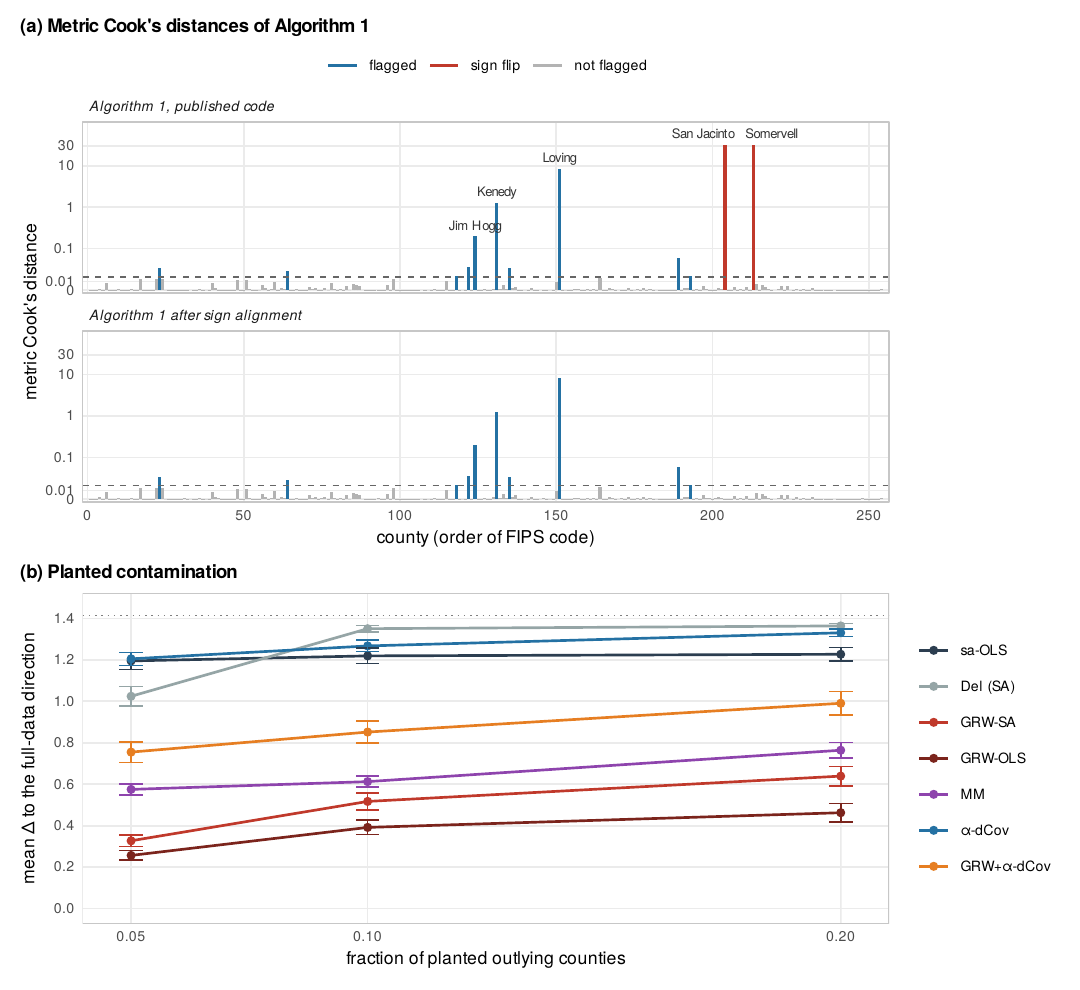}
\caption{Texas COVID-19 data. (a) Metric Cook's distances of Algorithm~1 of \citet{soale2025} with the published code and after aligning the sign of the leave-one-out MDS coordinate; red: counties flagged only because the recomputed coordinate is the mirror image of the full-sample one; dashed: cut-off $4/(n-p-1)$. (b) Mean $\Delta$ between the direction estimated after planting bad leverage points and the method's own full-data direction, with $\pm2$ standard errors; dotted: $\sqrt2$.}
\label{fig:texas}
\end{figure}

With the published code, Algorithm~1 of \citet{soale2025} flags 12 counties, two of them --- San Jacinto and Somervell --- only because the MDS coordinate recomputed without the county is the mirror image of the full-sample one (Figure~\ref{fig:texas}(a)): their Cook's distances are about 31.4, and below $10^{-3}$ once the sign is aligned. Which counties are flagged in this way can depend on the computer and its numerical library, and on our computer the same code does not reproduce Table~6 of \citet{soale2025} exactly (Supplement, Section~\ref{S-sec:texas}). Four counties named in that paper --- Falls, Hockley, Moore and Orange --- have Cook's distances between $9.9\times10^{-5}$ and $2.4\times10^{-3}$ here, far below the cut-off, so the same mirroring may explain why they were flagged there. After alignment, ten counties remain, the same as with the corrected implementation. GRW gives weight zero to 45 counties, including nine of these ten, mostly small rural counties with few reports and many zero rates (Table~\ref{S-tab:S-texas-counties}); the GRW directions describe the other 209 counties.

Table~\ref{tab:texas} summarizes the results. Since the true direction is unknown, we evaluate the directions by 10-fold cross-validated prediction of the county distributions and by their stability. No method predicts much better than the predictor-free reference (median $W_2$ error 257.2): the other methods are close to it (257.0--263.6), and the classical sa-OLS and FOLS are worse (273.5 and 283.0). When 5--20\% of the counties are replaced by bad leverage points, GRW-OLS and GRW-SA stay closest to their full-data directions (mean $\Delta$ 0.26--0.64), whereas the MM baseline moves by 0.58--0.76, unweighted FOLS by 0.87--1.01, and deletion, sa-OLS and $\alpha$-dCov by 0.96--1.37, at 10--20\% about as far as a direction drawn uniformly at random (1.32 for $p=8$; Figure~\ref{fig:texas}(b)); GRW-FOLS (0.78--1.07) and GRW+$\alpha$-dCov (0.76--0.99) are not protected in this sense. This criterion measures how far the planted points move a method, not whether its direction is correct, and the directions vary considerably even without contamination: under bootstrap resampling GRW-FOLS and GRW+$\alpha$-dCov vary nearly as much as a random direction (mean $\Delta$ 1.08 and 1.07, against 1.32), and GRW-SA more than sa-OLS (1.00 against 0.68), which reflects the weak signal in these data. In short, the directions of GRW-SA and GRW-OLS move least under planted contamination, with prediction errors close to those of the other methods. The entries of the published deletion, of $\alpha$-dCov and of GRW+$\alpha$-dCov in Table~\ref{tab:texas}, and hence the in-sample error $\delta$ of \citet{soale2025}, depend on the computer through the mirroring and the local optimizer (Supplement, Section~\ref{S-sec:texas}); those of the other methods do not.

\begin{table}[!htbp]\centering\scriptsize\setlength{\tabcolsep}{2.4pt}
\caption{Texas COVID-19 data. Number of counties removed (deletion) or with weight zero (GRW); in-sample prediction error $\delta$ of \citet{soale2025}, computed as in Table~6 of that paper on the counties not flagged by its Algorithm~1; 10-fold cross-validated median $W_2$ prediction error (20 repetitions; predictor-free reference 257.2) and percentage of test counties predicted better than by the reference; mean $\Delta$ to the method's own full-data direction with 5, 10 and 20\% planted bad leverage points (100 data sets each) and over 200 bootstrap samples (standard errors at most 0.04). Best value of each row except $\delta$ in bold. Columns as in Table~\ref{tab:summary}.}\label{tab:texas}
\begin{tabular}{@{}l*{12}{c}@{}}\toprule
& \multicolumn{4}{c}{sa-OLS family} & \multicolumn{4}{c}{FOLS family} & \multicolumn{4}{c}{other}\\\cmidrule(lr){2-5}\cmidrule(lr){6-9}\cmidrule(l){10-13}
& classical & Del-pub & Del & GRW & classical & Del-pub & Del & GRW & GRW-OLS & MM & \(\alpha\)-dCov & \shortstack{GRW+\\\(\alpha\)-dCov}\\
\midrule
\shortstack[l]{removed (Del) /\\ weight 0 (GRW)} & 0 & 12 & 10 & 45 & 0 & 12 & 10 & 45 & 45 & 0 & 0 & 45\\
in-sample $\delta$ & 832.1 & 827.3 & 827.2 & 830.1 & 832.8 & 828.4 & 828.3 & 831.6 & 831.7 & 832.0 & 828.4 & 832.9\\
CV median $W_2$ error & 273.5 & 262.4 & 263.6 & 257.9 & 283.0 & 261.1 & 260.8 & 257.6 & \textbf{257.0} & 257.8 & 257.3 & 260.2\\
\% better than ref. & 44 & 52 & 52 & \textbf{59} & 40 & 47 & 48 & 52 & 57 & 51 & 49 & 46\\
\midrule
planted 5\% & 1.19 & 0.96 & 1.02 & 0.33 & 0.87 & 1.00 & 1.09 & 0.78 & \textbf{0.26} & 0.58 & 1.20 & 0.76\\
planted 10\% & 1.22 & 1.35 & 1.35 & 0.52 & 0.97 & 1.36 & 1.37 & 1.03 & \textbf{0.39} & 0.61 & 1.27 & 0.85\\
planted 20\% & 1.23 & 1.37 & 1.36 & 0.64 & 1.01 & 1.36 & 1.36 & 1.07 & \textbf{0.46} & 0.76 & 1.33 & 0.99\\
bootstrap & 0.68 & 0.82 & 0.79 & 1.00 & 0.75 & 1.07 & 1.10 & 1.08 & 0.81 & \textbf{0.65} & 1.10 & 1.07\\
\bottomrule\end{tabular}\end{table}

\FloatBarrier
\subsection{Physical Activity Distributions in NHANES}\label{subsec:nhanes}

The second application uses the accelerometry data of NHANES 2011--2014, for which \citet{ghosal2023nhanes} fitted a partially linear single-index Fr\'echet model to distributional profiles of physical activity. We use the minute-level wrist accelerometry files of \citet{koffman2026nhanes}, in monitor-independent movement summary units \citep[MIMS;][]{john2019mims}. The response of each participant is the distribution of the MIMS values over the minutes of wake wear time on days with at least 600 such minutes, for participants with at least four such days, represented by its quantile function on 100 probabilities, so that $W_2$ is the $L^2$ distance between quantile functions. The $p=7$ predictors are age, body-mass index, waist circumference, systolic and diastolic blood pressure, HbA1c and total cholesterol \citep{cdc2014nhanes}; 7672 participants aged 20 to 80 have complete data. Because Algorithm~1 of \citet{soale2025} refits the MDS once per case, we analyze a random sample of $n=1000$ participants and use the other 6672 as an external test set; the methods and distances are those of Section~\ref{subsec:texas} (details in Section~\ref{S-sec:nhanes} of the Supplement). We ignore the survey weights and the complex sampling design, so that the results describe the sample rather than the US population. These data also fit the setting: each participant contributes a whole distribution, the predictors are continuous clinical measurements, and participants with unusually high or low activity give responses far from the others, which the isolation factor is designed to screen; the large sample also allows an external test set.

\begin{figure}[!htbp]\centering
\includegraphics[width=\linewidth]{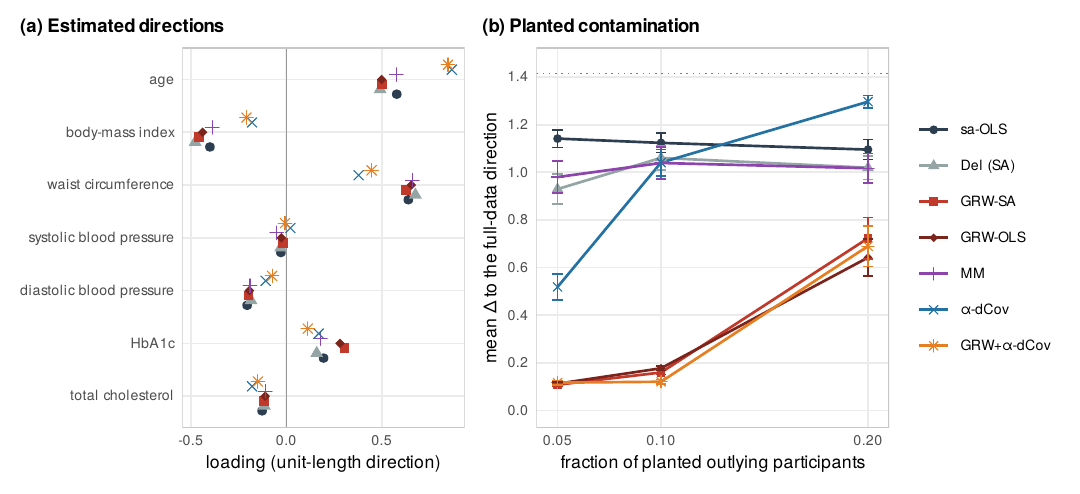}
\caption{NHANES data. (a) Estimated directions on the analysis sample (unit length, standardized predictors, signs aligned with sa-OLS). (b) Mean $\Delta$ between the direction estimated after planting bad leverage points and the method's own full-data direction, with $\pm2$ standard errors; dotted: $\sqrt2$.}
\label{fig:nhanes}
\end{figure}

In contrast to the Texas data, the MDS coordinate recomputed by Algorithm~1 of \citet{soale2025} is never mirrored here, and the published and corrected implementations flag 42 and 48 participants, 42 of them in common. GRW gives weight zero to 83 participants: 62 through the isolation factor, all of them participants whose activity is unusually high or low (42 above the 95th and 20 below the 5th percentile of the mean MIMS), and 21 with high leverage (mean HbA1c 8.5 and body-mass index 32.7, against 5.7 and 28.8 for the others). All directions except the two based on $\alpha$-dCov are similar (pairwise $\Delta$ between 0.03 and 0.27; Figure~\ref{fig:nhanes}(a)). The index loads positively on waist circumference and age and negatively on body-mass index, two strongly correlated predictors ($r=0.90$), so that it can be read as waist circumference relative to body-mass index, that is, central adiposity. The directions of $\alpha$-dCov and GRW+$\alpha$-dCov give most weight to age; their $\Delta$ to each of the ten other directions is at least 0.62 and 0.54, respectively. These two directions change slightly with the computer, through the local optimizer of $\alpha$-dCov, but remain more than $0.5$ from the others.

Table~\ref{tab:nhanes} summarizes the results. Here the predictors carry information: every method predicts the activity distributions of the external participants better than the predictor-free reference (median $W_2$ error 2.76--2.78 against 3.05) and is closer to the observed distribution than the reference for about 59\% of them, but the differences between methods are below 1\% of the reference. When 5\% or 10\% of the analysis sample are replaced by bad leverage points, the GRW estimators remain close to their full-data directions, that is, their directions on the uncontaminated analysis sample (mean $\Delta$ 0.11--0.18), whereas the classical estimators, deletion and the MM baseline lose them (0.89--1.14), and $\alpha$-dCov does so from 10\% on; at 20\%, GRW is also affected (0.64--0.72), but less than all competitors (0.99--1.30; Figure~\ref{fig:nhanes}(b)). Without contamination, the GRW directions vary more than those of unweighted sa-OLS and FOLS between bootstrap samples (mean $\Delta$ 0.32--0.35 against 0.25--0.26) and between five disjoint samples of 1000 participants (0.57--0.70 against 0.30--0.40), mainly in the split between the collinear predictors body-mass index and waist circumference and in the loading of HbA1c.

\begin{table}[!htbp]\centering\scriptsize\setlength{\tabcolsep}{2.4pt}
\caption{NHANES 2011--2014 physical-activity distributions ($n=1000$, $p=7$). Participants removed (deletion) or with weight zero (GRW); median $W_2$ prediction error for the 6672 participants outside the analysis sample (predictor-free reference 3.05) and percentage of them predicted better than by the reference; 10-fold cross-validated median $W_2$ error (20 repetitions; reference 3.05); mean $\Delta$ to the method's own full-data direction with 5, 10 and 20\% planted bad leverage points (100 data sets each; standard errors at most 0.04), over 200 bootstrap samples, and between the directions of 5 disjoint random samples of size 1000. The published deletion is computed only on the full data. Best value of each row in bold. Columns as in Table~\ref{tab:summary}.}\label{tab:nhanes}
\begin{tabular}{@{}l*{12}{c}@{}}\toprule
& \multicolumn{4}{c}{sa-OLS family} & \multicolumn{4}{c}{FOLS family} & \multicolumn{4}{c}{other}\\\cmidrule(lr){2-5}\cmidrule(lr){6-9}\cmidrule(l){10-13}
& classical & Del-pub & Del & GRW & classical & Del-pub & Del & GRW & GRW-OLS & MM & \(\alpha\)-dCov & \shortstack{GRW+\\\(\alpha\)-dCov}\\
\midrule
\shortstack[l]{removed (Del) /\\ weight 0 (GRW)} & 0 & 42 & 48 & 83 & 0 & 42 & 48 & 83 & 83 & 0 & 0 & 83\\
external median $W_2$ & 2.782 & 2.775 & 2.777 & 2.773 & 2.775 & 2.772 & 2.767 & 2.768 & 2.769 & 2.769 & 2.775 & \textbf{2.759}\\
\% better than ref. & 59.0 & 58.9 & 59.0 & 59.2 & \textbf{59.4} & 59.3 & 59.4 & 59.4 & 59.2 & 59.3 & 58.9 & 59.1\\
CV median $W_2$ & 2.791 & -- & 2.796 & 2.779 & 2.801 & -- & 2.799 & 2.785 & 2.785 & 2.789 & 2.755 & \textbf{2.752}\\
\midrule
planted 5\% & 1.14 & -- & 0.93 & 0.11 & 1.10 & -- & 0.89 & \textbf{0.11} & 0.11 & 0.98 & 0.52 & 0.12\\
planted 10\% & 1.12 & -- & 1.06 & 0.16 & 1.10 & -- & 1.04 & 0.16 & 0.18 & 1.04 & 1.04 & \textbf{0.12}\\
planted 20\% & 1.10 & -- & 1.02 & 0.72 & 1.04 & -- & 0.99 & 0.65 & \textbf{0.64} & 1.02 & 1.30 & 0.69\\
bootstrap & 0.26 & -- & 0.27 & 0.35 & 0.25 & -- & \textbf{0.25} & 0.32 & 0.35 & 0.29 & 0.25 & 0.30\\
disjoint samples & 0.40 & -- & 0.42 & 0.70 & \textbf{0.30} & -- & 0.34 & 0.57 & 0.67 & 0.50 & 0.44 & 0.51\\
\bottomrule\end{tabular}\end{table}

\FloatBarrier

\section{Discussion}\label{sec:discussion}
GRW turns the screening of influential observations in single-index Fr\'echet regression from a deletion rule into case weights that are computed once and can be used with any estimator that accepts case weights, and as a screen with those that do not. Its formal guarantee concerns the inputs that receive positive weight: the response objects stay in a bounded region when at most $\lfloor n/2\rfloor-1$ of the cases are replaced, and cases with $rd_i\ge4q_x$ receive weight zero through the Mallows factor. In the simulations, GRW keeps the mean $\Delta$ below about 0.5 under scattered bad leverage points up to 30\% contamination and under tightly clustered ones up to 20\% (Section~\ref{subsec:simresults}), whereas deletion and $\alpha$-dCov become much less accurate as the contamination grows. GRW is also more accurate than $\alpha$-dCov under magnitude outliers (T1) in most settings. In the two applications its directions move less than those of the methods that do not use GRW under planted bad leverage points (in the Texas data for GRW-SA and GRW-OLS but not for GRW-FOLS). Keeping the isolation factor in the final weights prevents the readmission of objects that are far from all others but invisible in the two MDS axes; in the model constructed for this purpose the direction is lost without it, whereas in Models II--VII its effect is small but consistent (Section~\ref{subsec:ablation}).

The weights have a cost. GRW gives weight zero to 10--32\% of the cases in clean simulated data and to 8--18\% in the applications, and it treats good leverage points like bad ones; $\alpha$-dCov is therefore more accurate with shape outliers and good leverage points, and the GRW directions are somewhat more variable than the classical ones in clean data. In practice, $\alpha$-dCov alone is preferable when bad leverage points can be ruled out, and GRW, or $\alpha$-dCov on the cases retained by GRW, when they cannot; in the simulations the combination had the smallest average error in every contamination type. GRW does not remove the limitation of OLS-type estimators to links that are not even in the index, and the applications support its robustness but not an improvement in prediction. In our designs, in the simulations as well as in the planted contamination of the applications, the responses of the bad leverage points are also outlying, so that the response-space screening contributes to their removal; leverage points with typical responses were not studied. Natural next steps are a readmission rule that does not penalize leverage when the residual of a case is small, extensions to more than one index and tuning constants chosen by the data; on the theoretical side, the population form of the GRW weight, sample consistency, the influence function of the data-adaptive weights, the breakdown point of the direction and asymptotic distribution theory remain open.

\section*{Supplementary Materials}
\begin{description}
\item[PDF supplement:] Algorithm~\ref{S-alg:grw}, the proofs of the results in Section~\ref{sec:theory}, the invariance of the weights, additional simulation results for Section~\ref{sec:sims} and details of the applications in Section~\ref{sec:real}.
\item[R code and data:] Three archives with the R code, the data and the stored results that reproduce the simulation studies in Section~\ref{sec:sims} (GRW\_simulations) and the analyses of the Texas and NHANES data in Section~\ref{sec:real} (GRW\_texas, GRW\_nhanes).
\item[Data Availability Statement]
The Texas COVID-19 data are those of the supplementary material of \citet{soale2025}. The NHANES minute-level activity data are available from PhysioNet \citep{koffman2026nhanes}, and the covariates were obtained from the NHANES website \citep{cdc2014nhanes}; the prepared analysis data set and the scripts that build it from these public files are included in the code supplement. The R code and the stored results that reproduce all tables and figures are provided as online supplementary material; the archives document the software versions, the seeds and the numbers of cores used.
\end{description}

\section*{Disclosure Statement}
The author reports there are no competing interests to declare.

\section*{Declaration of generative AI and AI-assisted technologies in the writing process}
During the preparation of this work, the author used a large language model (Claude, Anthropic) to review and edit the author's analysis code and manuscript text. The author wrote the code and the manuscript, designed the study, specified every methodological decision, and verified the code and the results. The author takes full responsibility for the content of the published article.

\bibliographystyle{agsm}
\bibliography{refs_competitors}

@article{huang2026fdsdr,
  author  = {Huang, Hsin-Hsiung and Yu, Feng and Li, Kang and Zhang, Teng},
  title   = {Fr{\'e}chet Sufficient Dimension Reduction for Metric Space-Valued Data via Distance Covariance},
  journal = {Journal of Computational and Graphical Statistics},
  volume  = {35}, number = {3}, year = {2026},
  doi     = {10.1080/10618600.2025.2602521},
  note    = {arXiv:2412.13122}
}

@article{huang2024adcov,
  author  = {Huang, Hsin-Hsiung and Yu, Feng and Zhang, Teng},
  title   = {Robust Sufficient Dimension Reduction via $\alpha$-Distance Covariance},
  journal = {Journal of Nonparametric Statistics},
  volume  = {37}, number = {4}, pages = {796--811}, year = {2025},
  doi     = {10.1080/10485252.2024.2313137},
  note    = {arXiv:2402.00778}
}

@manual{rsdr2025,
  author = {Chen, Sheau-Chiann and Zhao, Shilin and Huang, Hsin-Hsiung},
  title  = {{rSDR}: Robust Sufficient Dimension Reduction},
  year   = {2025}, note = {R package version 1.0.3.0, CRAN}
}

@article{zhang2019bcov,
  author  = {Zhang, Jia and Chen, Xin},
  title   = {Robust Sufficient Dimension Reduction via Ball Covariance},
  journal = {Computational Statistics \& Data Analysis},
  volume  = {140}, pages = {144--154}, year = {2019}
}

@article{pan2020ball,
  author  = {Pan, Wenliang and Wang, Xueqin and Zhang, Heping and Zhu, Hongtu and Zhu, Jin},
  title   = {Ball Covariance: A Generic Measure of Dependence in {B}anach Space},
  journal = {Journal of the American Statistical Association},
  volume  = {115}, number = {529}, pages = {307--317}, year = {2020},
  doi     = {10.1080/01621459.2018.1543600}
}

@misc{leyder2025robust,
  author  = {Leyder, Sarah and Raymaekers, Jakob and Rousseeuw, Peter J.},
  title   = {Robust Distance Covariance},
  howpublished = {\emph{International Statistical Review}},
  year    = {2026},
  doi     = {10.1111/insr.70005},
  note    = {With discussion. Early View, doi:10.1111/insr.70005}
}

@misc{li2025robustfrechet,
  author = {Li, Hao and Sugasawa, Shonosuke and Katayama, Shota},
  title  = {Robust Global {F}r{\'e}chet Regression via Weight Regularization},
  year   = {2025}, note = {arXiv:2511.03694}
}

@article{soale2025,
  author  = {Soale, Abdul-Nasah},
  title   = {Detecting Influential Observations in Single-Index {F}r{\'e}chet Regression},
  journal = {Technometrics}, volume = {67}, number = {2}, pages = {311--322}, year = {2025},
  doi     = {10.1080/00401706.2024.2441683}
}

@misc{soale2023sdr,
  author = {Soale, Abdul-Nasah and Dong, Yuexiao},
  title  = {Sufficient Dimension Reduction for Regression with Metric Space-Valued Responses},
  year   = {2023}, note = {arXiv:2310.12402}
}

@article{zhang2024frechet,
  author  = {Zhang, Qi and Xue, Lingzhou and Li, Bing},
  title   = {Dimension Reduction for {F}r{\'e}chet Regression},
  journal = {Journal of the American Statistical Association},
  volume  = {119}, number = {548}, pages = {2733--2747}, year = {2024},
  doi     = {10.1080/01621459.2023.2277406}
}

@article{ying2025dwdr,
  author  = {Ying, Chao and Yu, Zhou and Zhang, Xin},
  title   = {Distance Weighted Directional Regression for {F}r{\'e}chet Sufficient Dimension Reduction},
  journal = {Biometrics}, volume = {81}, number = {2}, pages = {ujaf051}, year = {2025}
}

@article{petersen2019frechet,
  author  = {Petersen, Alexander and M{\"u}ller, Hans-Georg},
  title   = {Fr{\'e}chet Regression for Random Objects with {E}uclidean Predictors},
  journal = {The Annals of Statistics}, volume = {47}, number = {2}, pages = {691--719}, year = {2019}
}

@article{rvd1999,
  author  = {Rousseeuw, Peter J. and Van Driessen, Katrien},
  title   = {A Fast Algorithm for the Minimum Covariance Determinant Estimator},
  journal = {Technometrics}, volume = {41}, number = {3}, pages = {212--223}, year = {1999}
}

@article{coakley1993bounded,
  author  = {Coakley, Clint W. and Hettmansperger, Thomas P.},
  title   = {A Bounded Influence, High Breakdown, Efficient Regression Estimator},
  journal = {Journal of the American Statistical Association}, volume = {88}, number = {423}, pages = {872--880}, year = {1993}
}

@article{gervini2002class,
  author  = {Gervini, Daniel and Yohai, V{\'i}ctor J.},
  title   = {A Class of Robust and Fully Efficient Regression Estimators},
  journal = {The Annals of Statistics}, volume = {30}, number = {2}, pages = {583--616}, year = {2002}
}

@article{bhattacharjee2023single,
  author  = {Bhattacharjee, Satarupa and M{\"u}ller, Hans-Georg},
  title   = {Single Index {F}r{\'e}chet Regression},
  journal = {The Annals of Statistics}, volume = {51}, number = {4}, pages = {1770--1798}, year = {2023},
  doi     = {10.1214/23-AOS2307}
}

@article{ghosal2023frechet,
  author  = {Ghosal, Aritra and Meiring, Wendy and Petersen, Alexander},
  title   = {{F}r{\'e}chet Single Index Models for Object Response Regression},
  journal = {Electronic Journal of Statistics}, volume = {17}, number = {1}, pages = {1074--1112}, year = {2023},
  doi     = {10.1214/23-EJS2120}
}

@inproceedings{cui2026deep,
  author    = {Cui, Muqing and Zhou, Yidong and Iao, Su I and M{\"u}ller, Hans-Georg},
  title     = {Deep Single-Index {F}r{\'e}chet Regression},
  booktitle = {Proceedings of the 43rd International Conference on Machine Learning},
  series    = {PMLR}, volume = {306}, pages = {22095--22114}, year = {2026}
}

@article{ying2022frechet,
  author  = {Ying, Chao and Yu, Zhou},
  title   = {{F}r{\'e}chet Sufficient Dimension Reduction for Random Objects},
  journal = {Biometrika}, volume = {109}, number = {4}, pages = {975--992}, year = {2022}
}

@article{dong2022fksir,
  author  = {Dong, Yushen and Wu, Yichao},
  title   = {{F}r{\'e}chet Kernel Sliced Inverse Regression},
  journal = {Journal of Multivariate Analysis}, volume = {191}, pages = {105032}, year = {2022},
  doi     = {10.1016/j.jmva.2022.105032}
}

@article{virta2022sir,
  author  = {Virta, Joni and Lee, Kuang-Yao and Li, Lexin},
  title   = {Sliced Inverse Regression in Metric Spaces},
  journal = {Statistica Sinica}, volume = {32}, pages = {2315--2337}, year = {2022}
}

@article{huang2026knn,
  author  = {Huang, Xueyan and Qiu, Rui and Yu, Zhou},
  title   = {{F}r{\'e}chet k{NN}-Based Sufficient Dimension Reduction},
  journal = {Journal of Multivariate Analysis}, volume = {212}, pages = {105566}, year = {2026}
}

@inproceedings{weng2024sparse,
  author    = {Weng, Jiaying and Ke, Chenlu and Wang, Pei},
  title     = {Sparse {F}r{\'e}chet Sufficient Dimension Reduction via Nonconvex Optimization},
  booktitle = {Conference on Parsimony and Learning (CPAL)}, series = {PMLR}, volume = {234}, pages = {39--53}, year = {2024}
}

@article{chen2026intrinsic,
  author  = {Chen, Baiyu and Dai, Shuang and Yu, Zhou},
  title   = {Intrinsic Minimum Average Variance Estimation for Dimension Reduction with Symmetric Positive-Definite Matrices and Beyond},
  journal = {Statistica Sinica}, volume = {36}, number = {3}, pages = {1133--1153}, year = {2026},
  doi     = {10.5705/ss.202023.0268}
}

@misc{soale2026metric,
  author  = {Soale, Abdul-Nasah and Ma, Congli and Chen, Siyu and Koomson, Obed},
  title   = {On Metric Choice in Dimension Reduction for {F}r{\'e}chet Regression},
  howpublished = {\emph{International Statistical Review}}, year = {2026},
  doi     = {10.1111/insr.12615}, note = {arXiv:2410.01783}
}

@techreport{becker1998robust,
  author      = {Becker, Claudia and Gather, Ursula and Hilker, Torsten},
  title       = {Robust Sliced Inverse Regression Procedures},
  institution = {SFB 475, Universit{\"a}t Dortmund}, number = {22/98}, year = {1998}
}

@article{prendergast2005influence,
  author  = {Prendergast, Luke A.},
  title   = {Influence Functions for Sliced Inverse Regression},
  journal = {Scandinavian Journal of Statistics}, volume = {32}, number = {3}, pages = {385--404}, year = {2005},
  doi     = {10.1111/j.1467-9469.2005.00447.x}
}

@phdthesis{genschel2005robustness,
  author = {Genschel, Ulrike},
  title  = {Robustness Concepts for Sliced Inverse Regression},
  school = {Universit{\"a}t Dortmund}, year = {2005}
}

@article{dong2015robust,
  author  = {Dong, Yuexiao and Yu, Zhou and Zhu, Liping},
  title   = {Robust Inverse Regression for Dimension Reduction},
  journal = {Journal of Multivariate Analysis}, volume = {134}, pages = {71--81}, year = {2015}
}

@article{iaci2025robust,
  author  = {Iaci, Ross and Sriram, T. N.},
  title   = {Robust Recovery of the Central Subspace for Regression Using the Influence Function of the {R}{\'e}nyi Divergence},
  journal = {Statistica Sinica}, volume = {35}, number = {2}, pages = {799--830}, year = {2025},
  doi     = {10.5705/ss.202022.0351}
}

@article{li2026robustlocal,
  author  = {Li, Yongjiang and Zhu, Linlin},
  title   = {Robust Local {W}asserstein {F}r{\'e}chet Regression under Admissible Transport Deformations},
  journal = {Statistical Papers}, volume = {67}, pages = {124}, year = {2026},
  doi     = {10.1007/s00362-026-01905-z}
}

@inproceedings{kim2025frechetmedian,
  author    = {Kim, Jakwang and Park, Jiyoung and Bhattacharya, Anirban},
  title     = {Robust Estimation in Metric Spaces: Achieving Exponential Concentration with a {F}r{\'e}chet Median},
  booktitle = {Proceedings of the 28th International Conference on Artificial Intelligence and Statistics (AISTATS)},
  series    = {PMLR}, volume = {258}, pages = {4222--4230}, year = {2025}
}

@article{dai2023tukey,
  author  = {Dai, Xiongtao and Lopez-Pintado, Sara},
  title   = {Tukey's Depth for Object Data},
  journal = {Journal of the American Statistical Association}, volume = {118}, number = {543}, pages = {1760--1772}, year = {2023},
  doi     = {10.1080/01621459.2021.2011298}
}

@misc{virta2023spatial,
  author  = {Virta, Joni},
  title   = {Spatial Depth for Data in Metric Spaces},
  howpublished = {\emph{Scandinavian Journal of Statistics}}, year = {2026},
  doi     = {10.1111/sjos.70054}, note = {Early View, doi:10.1111/sjos.70054}
}

@article{yohai1987high,
  author  = {Yohai, V{\'i}ctor J.},
  title   = {High Breakdown-Point and High Efficiency Robust Estimates for Regression},
  journal = {The Annals of Statistics}, volume = {15}, number = {2}, pages = {642--656}, year = {1987}
}

@article{cook1994reweighting,
  author  = {Cook, R. Dennis and Nachtsheim, Christopher J.},
  title   = {Reweighting to Achieve Elliptically Contoured Covariates in Regression},
  journal = {Journal of the American Statistical Association}, volume = {89}, number = {426}, pages = {592--599}, year = {1994}
}

@article{li1989regression,
  author  = {Li, Ker-Chau and Duan, Naihua},
  title   = {Regression Analysis under Link Violation},
  journal = {The Annals of Statistics}, volume = {17}, number = {3}, pages = {1009--1052}, year = {1989}
}

@misc{cdc2023covid,
  author = {{Centers for Disease Control and Prevention}},
  title  = {{United States} {COVID}-19 County Level of Community Transmission Historical Changes},
  year   = {2023}, howpublished = {\url{https://data.cdc.gov}},
  note   = {Accessed 13 March 2023}
}

@misc{census2022acs,
  author = {{U.S. Census Bureau}},
  title  = {{American Community Survey} 5-Year Estimates, 2020},
  year   = {2022}, howpublished = {\url{https://www.census.gov/data.html}},
  note   = {Accessed 26 October 2022}
}

@article{ghosal2023nhanes,
  author  = {Matabuena, Marcos and Ghosal, Aritra and Meiring, Wendy and Petersen, Alexander},
  title   = {Predicting Distributions of Physical Activity Profiles in the {N}ational {H}ealth and {N}utrition {E}xamination {S}urvey Database Using a Partially Linear {F}r{\'e}chet Single Index Model},
  journal = {Biostatistics}, volume = {26}, number = {1}, pages = {kxaf013}, year = {2025},
  doi     = {10.1093/biostatistics/kxaf013}
}

@article{john2019mims,
  author  = {John, Dinesh and Tang, Qu and Albinali, Fahd and Intille, Stephen},
  title   = {An Open-Source Monitor-Independent Movement Summary for Accelerometer Data Processing},
  journal = {Journal for the Measurement of Physical Behaviour}, volume = {2}, number = {4}, pages = {268--281}, year = {2019},
  doi     = {10.1123/jmpb.2018-0068}
}

@misc{koffman2026nhanes,
  author = {Koffman, Lily and Muschelli, John},
  title  = {Minute Level Step Counts and Physical Activity Data from the {National Health and Nutrition Examination Survey (NHANES)} 2011--2014},
  year   = {2026}, howpublished = {PhysioNet, version 1.0.2}, doi = {10.13026/d7sw-f662}
}

@misc{cdc2014nhanes,
  author = {{Centers for Disease Control and Prevention, National Center for Health Statistics}},
  title  = {{National Health and Nutrition Examination Survey} Data, 2011--2012 and 2013--2014},
  year   = {2014}, howpublished = {\url{https://wwwn.cdc.gov/nchs/nhanes/}}
}

@article{rousseeuw1984least,
  author  = {Rousseeuw, Peter J.},
  title   = {Least Median of Squares Regression},
  journal = {Journal of the American Statistical Association}, volume = {79}, number = {388}, pages = {871--880}, year = {1984}
}

@article{rousseeuw2006computing,
  author  = {Rousseeuw, Peter J. and Van Driessen, Katrien},
  title   = {Computing {LTS} Regression for Large Data Sets},
  journal = {Data Mining and Knowledge Discovery}, volume = {12}, number = {1}, pages = {29--45}, year = {2006}
}
\end{document}